\pdfoutput=1
\documentclass[a4paper,fleqn]{cas-dc}

\usepackage[numbers,sort&compress]{natbib}

\usepackage{algorithm}
\usepackage{algpseudocode}
\usepackage{breakcites}
\usepackage{float}
\usepackage{placeins}
\usepackage{tabularx}
\usepackage{tikz}
\usepackage[normalem]{ulem}
\usepackage{subcaption}
\usepackage{textcomp}

\newenvironment{varalgorithm}[1]
{\algorithm\renewcommand{\thealgorithm}{#1}}
{\endalgorithm}

\newcommand{\notationrow}[2]{%
  \noindent
  \begin{minipage}[t]{0.29\columnwidth}\raggedright #1
  \end{minipage}\hfill
  \begin{minipage}[t]{0.68\columnwidth}\raggedright #2
  \end{minipage}%
  \par\nobreak\vspace{0.5pt}%
  \noindent{\color{black!20}\rule{\columnwidth}{0.2pt}}%
\par\vspace{0.8pt}}

\newcommand{\notationheading}[1]{%
  \medskip\noindent\textbf{#1}\par\nobreak\vspace{1pt}%
  \noindent{\color{black!55}\rule{\columnwidth}{0.5pt}}%
\par\nobreak\vspace{2pt}}
\def\BibTeX{{\rm B\kern-.05em{\sc i\kern-.025em b}\kern-.08em
T\kern-.1667em\lower.7ex\hbox{E}\kern-.125emX}}

\def \Cap           {\textsc{cap}}
\def \CPU           {\textsc{cpu}}
\def \CUIdx         {c}
\def \CUSet         {C}
\def \CU            {\ensuremath{\mbox{\upshape\textsc{cu}}}}

\def \DU                {\ensuremath{\mbox{\upshape\textsc{du}}}}
\def \DUIdx             {d}
\def \DUSet             {D}

\def \EnergyLink     {\epsilon_{\ServerIdx\ServerIdx'}^{\textsc{TX}}}

\def \idle          {\textsc{idle}}
\def \FlowIdx       {k}
\def \FlowSet       {K}
\def \IO            {\textsc{io}}
\def \PiwiseSet     {\mathcal{I}}

\def \LatencyBetweenServers {\delta_{\ServerIdx \ServerIdx'}}

\def \MaxUtil       {\rho_{\max}}
\def \MigrationIdx  {m}
\def \Migration     {\textsc{migr}}
\def \MigrationCoeff {\kappa}
\def \MaxCUPerDU    {\tau^{\max}_{\DU}}

\def \NameHeuristic {H\_EJPM}

\def \Temperature {\theta_\ServerIdx}

\def \Placement {\textsc{PLA}}
\def \PbName    {EJPM}

\def \PreviousPlacement {p}
\def \PreviousPlacementDU   {\PreviousPlacement_{\DUIdx \ServerIdx}^{\DU}}
\def \PreviousPlacementCU   {\PreviousPlacement_{\CUIdx \ServerIdx}^{\CU}}
\def \PreviousActive {\bar{a}}

\def \AssignCU {q}

\def \RAM                   {\textsc{ram}}
\def \ResourceIdx           {r}

\def \ResourceSet           {R}

\def \ServerIdx         {s}
\def \ServerSet         {S}

\def \SliceIdx          {\sigma}
\def \SliceSet          {S^{\textsc{sli}}}

\def \TrafficSet       {T}
\def \TrafficFlowData  {\TrafficSet}
\def \ControlInterval  {\Delta t}

\def \SliceSetLegacy   {S^{\textsc{sli}\_\textsc{leg}}}

\def \PiwiseIdx    {i}
\def \PiwisePower  {P^{\mathrm{inc}}}
\def \Utilization   {u}

\def \VarActive     {a}
\def \WakeUp        {w}

\begin{document}
\let\WriteBookmarks\relax
\def\floatpagepagefraction{1}
\def\textpagefraction{.001}

\shorttitle{Energy-aware Placement and Migration in O-RAN Edge Clouds}
\shortauthors{Tran et al.}

\title[mode=title]{Next-generation O-RAN Edge: Energy-aware Joint Placement and Migration of Cloud-Native Functions}

\author[1]{Nguyen Phuc Tran}[
  orcid=0000-0002-9792-7907
]
\author[1]{Brigitte Jaumard}[
  orcid=0000-0003-3443-4918
]
\cormark[1]
\ead{brigitte.jaumard@concordia.ca}
\author[2]{Oscar Delgado}

\affiliation[1]{
  organization={Department of Computer Science and Software Engineering, Concordia University},
  city={Montréal},
  state={Québec},
  country={Canada}
}

\affiliation[2]{
  organization={Ericsson},
  city={Montréal},
  state={Québec},
  country={Canada}
}

\cortext[1]{Corresponding author}

\begin{abstract}
  The transition toward Open Radio Access Networks (O-RANs) is reshaping how cellular infrastructure is deployed, managed, and optimized.
  This paper investigates the energy-aware joint placement and migration of cloud-native functions (CNFs) in an O-RAN edge cloud.
  We consider both a Single-CU-UP association model and a slice-aware Multi-CU-UP relaxation, in which distinct slice-flow groups of the same distributed unit (DU) may be assigned to different Centralized Unit User Plane (CU-UP) processing targets under one Centralized Unit Control Plane (CU-CP).
  For brevity, these scenarios are referred to as Single-CU and Multi-CU, respectively; Multi-CU never denotes multiple CU-CP associations.
  We formulate the problem as a Mixed-Integer Linear Program (MILP) that minimizes server, transmission, wake-up, and migration energy while satisfying server-resource capacities and one-way delay requirements over the F1 user-plane interface (F1-U) between each DU and its selected CU-UP in a fat-tree edge data center.
  To improve computational scalability, we also develop a deterministic $k$-means-based heuristic that approximates the MILP decisions without requiring repeated exact optimization.

  Over the evaluated 24-hour workload, the theoretical Multi-CU relaxation reduces modeled energy consumption by 5.7\% relative to the Single-CU baseline.
  For the Multi-CU case, the proposed heuristic remains within approximately 9.7\% of the proposed MILP, demonstrating a favorable trade-off between energy efficiency and computational tractability.
\end{abstract}

\begin{highlights}
\item A MILP jointly optimizes CNF placement, migration, and energy use.
\item Flexible DU-to-CU-UP associations are evaluated under a fixed CU-CP.
\item Single-CU-UP and slice-aware Multi-CU-UP operation are compared.
\item Multi-CU-UP lowers modeled MILP energy by 5.7\% over 24 hours.
\item The proposed heuristic's 24-hour Multi-CU energy is 9.7\% above the MILP; it is also compared with common cloud-placement heuristics.
\end{highlights}

\begin{keywords}
  Open Radio Access Network (O-RAN)
  \sep Cloud-native function placement and migration
  \sep Energy efficiency
  \sep DU--CU-UP connectivity
  \sep Edge computing
  \sep Network slicing
  \sep Green networking
  \sep O-Cloud workload management
\end{keywords}

\maketitle
\sloppy

\section{Introduction}

The transition to the O-RAN architecture is reshaping modern telecom networks and supporting the evolution from Fourth-Generation/Long-Term Evolution (4G/LTE) systems to 5G and B5G deployments.
Traditionally, Radio Access Networks (RANs) were implemented using proprietary, tightly integrated hardware stacks, where the radio, baseband processing, and control logic were co-located within a single cell-site enclosure, limiting deployment flexibility and scalability.
This monolithic RAN architecture, although stable, constrained innovation, enforced vendor lock-in, and restricted resource scaling capabilities.
The O-RAN Alliance defines a disaggregated architecture that splits the traditional RAN into three distinct functional entities~\cite{dai2024ran}.
This functional split (typically Split 7.2x) enables the DU and CU to be virtualized as Virtualized Network Functions (VNFs) or containerized as CNFs and deployed on Commercial Off-The-Shelf (COTS) servers within edge clouds or regional data centers.
The F1 interface comprises the F1 user-plane interface (F1-U) between a DU and a CU-UP and the F1 control-plane interface (F1-C) between a DU and its CU-CP~\cite{3gpp38401}.
In this paper, we consider O-RAN because it provides the open, disaggregated deployment and management context in which DU and CU functions can run as cloud-native workloads on an O-Cloud and be orchestrated by open management mechanisms. The DU-CU functional split considered here remains rooted in 3GPP NR; O-RAN motivates the corresponding cloud-native placement, migration, and service-management problem.
This disaggregation gives operators greater flexibility in associating and placing DU and CU functions across available edge-cloud resources~\cite{hojeij2024flexible}.
Consequently, these functions are not tied to fixed servers and can be relocated in response to changes in traffic demand, resource availability, or reliability requirements~\cite{di2022optimization}.
This flexibility, however, introduces the challenge of jointly determining placement and migration decisions while satisfying operational constraints.

\begin{figure}[pos=t]
  \centering
  \includegraphics[width=0.95\linewidth]{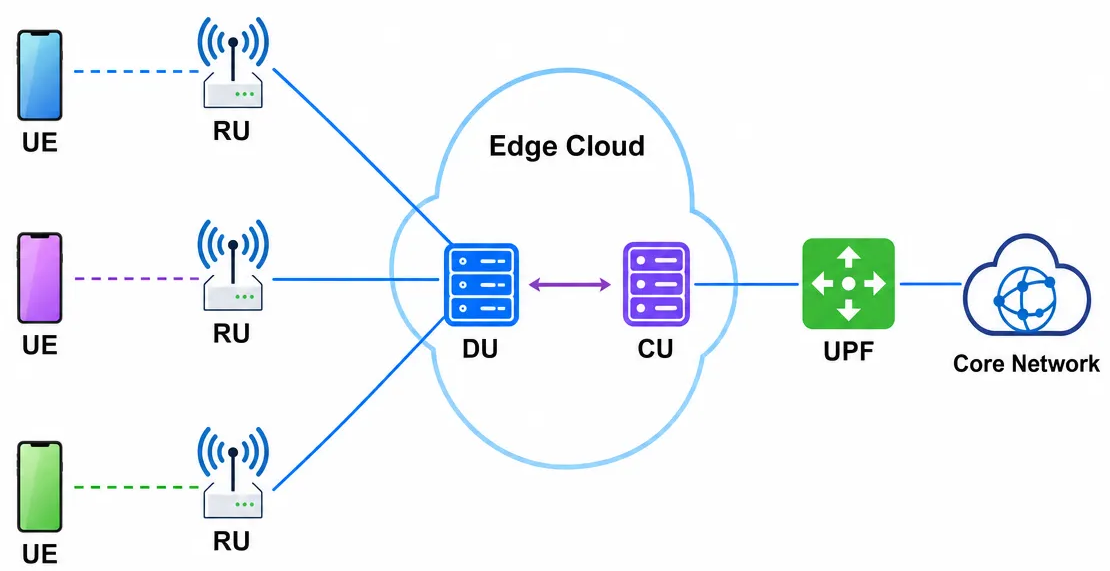}
  \caption{Scope of the proposed optimization. The model starts at the edge-cloud ingress and optimizes DU/CU-UP CNF placement and migration within one edge cloud.}
  \label{fig:overview_edge_cloud_scope}
\end{figure}

Edge computing further accelerates this transition by placing compute and storage resources closer to end users, thereby reducing delay and enhancing quality of service.
However, deciding where to place DUs and CU-UPs within a single edge cloud and when to migrate them requires careful planning.
Migration is needed because edge-cloud demand is time varying: low-demand periods create opportunities to consolidate CNFs and switch lightly loaded servers to idle or sleep states, whereas traffic hotspots can make a previously feasible placement violate resource-capacity or DU-to-CU-UP/F1-U-delay constraints. By relocating selected DU/CU-UP CNFs only when beneficial or necessary, the orchestrator can preserve feasibility under load fluctuations while reducing the number of active servers over time.
As illustrated in Fig.~\ref{fig:overview_edge_cloud_scope}, the model begins at the ingress of one selected edge cloud and optimizes DU/CU-UP placement and migration only within that cloud.
RU location, RU-DU transport and latency, and inter-edge-cloud relocation are external deployment and admission concerns; within the selected cloud, the model enforces resource and DU-to-CU-UP/F1-U delay constraints.

Energy consumption is also becoming a major concern in 5G and beyond.
Data centers, including edge clouds, consume nearly 1.7\% of global electricity, and their energy demand is projected to rise \cite{takci2025data}.
Edge environments add complexity because resources are geographically distributed, traffic patterns fluctuate continuously, and energy-aware placement becomes more challenging \cite{ismail2024powergen, 7393804}.
In this work, energy-aware placement jointly determines the locations of DU and CU-UP instances within one edge cloud and whether they should be migrated from their previous locations. The objective is to minimize active and idle server energy, intra-cloud transmission energy, wake-up energy, and migration overhead while satisfying server-resource capacities and one-way DU-to-CU-UP/F1-U delay bounds.
Most existing approaches prioritize load balancing or delay and do not jointly combine strict DU-to-CU-UP delay constraints~\cite{municio2023ran}, changing resource demand, and migration-aware placement.
Thus, when placement and migration are handled in isolation, the system misses opportunities to optimize energy consumption, delay, and overall performance.
A key architectural decision in next-generation RAN design is how DUs and CUs connect, since this choice determines how computation, fronthaul capacity, and delay constraints interact across the network~\cite{Trojer2021PacketFronthaul}.
In particular, the choice between forcing all slice-flow groups of a DU through one CU-UP processing target and allowing slice-specific groups to select different CU-UP processing targets fundamentally alters the feasible placement space.
The 3GPP CU-CP/CU-UP separation permits one DU to connect to multiple CU-UPs under one CU-CP, while a DU is normally connected to only one CU-CP~\cite{3gpp38401}.
Accordingly, Multi-CU is shorthand for Multi-CU-UP: the model may assign distinct slice-flow groups of a DU to different CU-UP processing targets, but each group has exactly one target and the DU's CU-CP association remains fixed.
This forward-looking relaxation abstracts implementation-specific user-plane steering and state-coordination mechanisms.
The purpose of this relaxation is not to claim immediate deployability, but to assess whether its potential energy savings are substantial enough to justify further architectural investigation.

To address these challenges, we propose an energy-aware framework for jointly optimizing CNF placement and migration under time-varying traffic demand.
The framework accounts for load-dependent incremental server power, fixed idle power, one-time server wake-up energy, transmission energy, physically calibrated migration energy, and DU-to-CU-UP/F1-U delay requirements.
To improve computational scalability, we further develop a deterministic $k$-means-based heuristic and evaluate it against the MILP model.
Our main contributions are as follows:

\begin{itemize}
  \item We formulate a MILP in which all power terms are converted to energy over each control interval, server wake-ups are represented by explicit off-to-on transition variables, and migration-energy coefficients are precomputed from the transferred state volume, path bandwidth, network energy intensity, and source and destination endpoint overheads.
  \item We introduce an explicit slice-flow-to-CU-UP assignment variable and a route variable indexed by the selected CU-UP, making Single-CU and Multi-CU transparent restrictions of the same formulation while retaining one fixed CU-CP association per DU.
  \item We specify an implementable four-phase heuristic with deterministic initialization, explicit candidate ordering, sequential migration acceptance, bounded feasibility repair, and no global completeness claim.
  \item We evaluate the formulation and heuristic under a time-varying 24-hour O-RAN workload and compare energy, delay feasibility, active servers, and migration behaviour with common placement baselines.
\end{itemize}


\section{Literature Review}

Generic VNF-placement studies established exact, metaheuristic, and hybrid methods for resource- and delay-constrained orchestration~\cite{ruiz2020genetic,yang2021delay,agarwal2018joint,golkarifard2021dynamic}. Optimization over time subsequently incorporated network-function relocation and reliability in 5G-RAN settings~\cite{di2022optimization}, while live-migration studies quantified the state-transfer and endpoint overheads that a physical migration-energy model must represent~\cite{ramanathan2021live}. These foundations motivate the rolling placement-and-migration structure used here, but they do not by themselves specify O-RAN DU--CU association.

Recent O-RAN research is more directly related to our work.
Murti~\textit{ et al.}~\cite{murti2024reconfigurations} use deep reinforcement learning to reconfigure functional splits, vCU/vDU locations, resource assignments, and routing under time-varying conditions while accounting for reconfiguration costs.
Hojeij~\textit{ et al.}~\cite{hojeij2023dynamic,hojeij2024flexible} study flexible O-DU/O-CU association and placement under changing resource bottlenecks.
Pires~\textit{ et al.}~\cite{pires2025optimizing} formulate energy-minimizing vRAN placement with flexible functional splits and transport-network energy.
Another study directly addresses energy-efficient placement and association in disaggregated O-RAN~\cite{hojeij2025energy}.
Sen~\textit{ et al.}~\cite{sen2025slice} and Mushtaq~\textit{ et al.}~\cite{mushtaq2023optimal} examine slice-aware functional splitting, placement, and routing.
Collectively, these studies establish energy-aware O-RAN placement as an active research area; accordingly, we do not claim energy minimization or DU/CU placement alone as novel.

Learning-based methods for O-RAN orchestration are also advancing rapidly.
The RFD-R framework dynamically repacks cloud-native RAN functions through merge, split, and move actions~\cite{sahin2026rfdr}.
CROWN applies cross-attention reinforcement learning to O-RAN control~\cite{monaco2026crown}.
These approaches target adaptation and scalability, but their learned policies and cost definitions differ from a solver-verifiable linear energy formulation.
By contrast, our MILP provides a transparent optimization reference, and the proposed deterministic heuristic approximates the decisions of the same formulation without acting as a learned controller.

Table~\ref{tab:related_work_positioning} positions this work against studies that address relevant subsets of the problem. Its distinguishing combination is rolling placement and migration, explicit per-slice-flow CU-UP assignment, a common-unit physical energy model, and a deterministic heuristic that approximates the same MILP decisions.

\begin{table*}[pos=t]
  \centering
  \caption{Comparison with related O-RAN orchestration studies.}
  \label{tab:related_work_positioning}
  \scriptsize
  \begin{tabularx}{\textwidth}{
      >{\raggedright\arraybackslash}p{0.1\textwidth}
      >{\raggedright\arraybackslash}X
      >{\centering\arraybackslash}p{0.07\textwidth}
      >{\centering\arraybackslash}p{0.11\textwidth}
      >{\centering\arraybackslash}p{0.12\textwidth}
    >{\centering\arraybackslash}p{0.17\textwidth}}
    \toprule
    \textbf{Work} & \textbf{Primary focus} & \textbf{Energy} & \makecell{\textbf{Migration/move}} & \makecell{\textbf{Slice-flow CU assign}} & \textbf{Solution approach} \\
    \midrule
    \cite{di2022optimization} & Reliable 5G-RAN optimization over time & Partial & Yes & No & \mbox{ILP + local search} \\
    \cite{murti2024reconfigurations} & Joint vRAN split, placement, and routing reconfiguration & No & Yes & No & \mbox{Action-branching D3QN} \\
    \cite{hojeij2023dynamic,hojeij2024flexible} & Flexible O-DU/O-CU association and placement & Partial & Partial & No & \mbox{ILP + RNN/heuristic} \\
    \cite{pires2025optimizing} & vRAN placement, splits, and transport energy & Yes & No & No & \mbox{MILP + heuristic} \\
    \cite{hojeij2025energy} & Energy-efficient O-RAN placement and association & Yes & No & No & \mbox{ILP + GNN} \\
    \cite{sen2025slice} & Slice-aware function splitting and placement & Yes & No & Partial & \mbox{MILP + heuristic} \\
    RFD-R~\cite{sahin2026rfdr} & AI-driven CNF repacking & Partial & Yes & No & \mbox{PPO-based DRL} \\
    \textbf{This work} & Rolling DU/CU-UP placement and migration & Yes & Yes & Yes & \mbox{MILP + heuristic} \\
    \bottomrule
  \end{tabularx}
  \vspace{1mm}
  \parbox{\textwidth}{\scriptsize\textit{Abbreviations:} D3QN: dueling double deep Q-network; DRL: deep reinforcement learning; GNN: graph neural network; ILP: integer linear programming; MILP: mixed-integer linear programming; PPO: proximal policy optimization; RNN: recurrent neural network.}
\end{table*}


\section{System Model and Assumptions}

The selected edge cloud is assumed to satisfy the fronthaul feasibility of its serving RUs before optimization. The model therefore starts from requests that have reached the edge-cloud ingress; RU-DU transport engineering and latency enforcement are outside its scope.
Traffic demand varies across optimization intervals but is deterministic within each interval.
At the beginning of interval $t$, the preprocessed demand for that interval is available as a known, fixed input that determines DU/CU-UP requirements for Central Processing Unit (CPU), Random Access Memory (RAM), and Input/Output (I/O) across heterogeneous slices such as enhanced Mobile Broadband (eMBB) and ultra-Reliable Low-Latency Communications (uRLLC).
Future demand, probability distributions, and stochastic scenarios are not used in the current-interval decision.
The edge cloud is homogeneous and adopts a fat-tree topology, where all interconnects use optical fiber links with a propagation speed of approximately $2 \cdot 10^8$ m/s; accordingly, propagation delay is modelled using fixed intra-pod values and negligible intra-server latency.
Although real-world congestion can introduce jitter, the current model treats link delays as static and additive.
All latency quantities constrained by the model are one-way delays.
The model considers only the intra-edge-cloud F1-U path between each DU and its selected CU-UP, comprising server-to-server delay within the fat-tree and user-plane processing delay at the DU and CU-UP endpoints.
RU-DU fronthaul propagation and jitter, User Equipment (UE) radio-access delay, CU-to-core transport delay, F1-C control-plane delay, and non-F1 application/core processing delay are outside the present scope and are left for future extensions.
Server power is modelled as fixed idle power plus a load-dependent piecewise-linear incremental component. Power in watts is converted to energy over the control interval, and every off-to-on transition incurs one fixed wake-up energy charge. This charge is an input parameter that must be calibrated for the deployed server platform or varied in sensitivity analysis~\cite{lin2020taxonomy}.
To maintain carrier-grade stability and provide a buffer for instantaneous traffic bursts, particularly for delay-critical uRLLC slices, we define the maximum operational capacity of each server at 70\%.
During preprocessing, traffic demand exceeding this threshold causes the assumed external scaling engine to generate additional CNF instances.
The proposed MILP and heuristic do not decide when or how to scale; they only place and migrate the resulting instances while respecting the 70\% operational-capacity limit.


\section{Energy-aware Joint Placement and Migration (\PbName) Problem Statement}

Given these assumptions, the joint CNF placement-and-migration problem minimizes edge-cloud energy consumption.
The constraints enforce \textit{(i)} server-resource capacity, \textit{(ii)} the scenario-specific one-way DU-to-CU-UP/F1-U delay budgets defined in Section~\ref{sec:traffic_setup}, and \textit{(iii)} the logical connectivity rules of the Single-CU and Multi-CU scenarios.

For each interval, traffic is represented by flows that may contain multiple slice-specific components; a flow adopts the most stringent delay requirement among its components.
We assume that traffic can be logically differentiated into per-slice flows at the DU level. Fine-grained assignment of these groups among CU-UP targets is treated as an abstract forward-looking relaxation, consistent with research on slice-aware splitting and placement~\cite{sen2025slice, mushtaq2023optimal}.
Throughout the formulation, the set $\CUSet$ and the term CU target refer to candidate CU-UP processing instances on the F1-U user-plane path.
The DU's single logical CU-CP/F1-C association is assumed to be established beforehand and remains fixed; CU-CP placement, migration, replication, and resiliency are outside the scope of the optimization.
This is a logical control-plane restriction, not a requirement that the CU-CP use dedicated physical hardware.
Each server may host multiple DU and CU-UP instances subject to resource and delay constraints; the model abstracts their container implementation and represents only their processing requirements.

Accordingly, the optimization selects DU and CU-UP placements for each slice flow so that all resource, delay, and connectivity constraints are satisfied.
The legacy traffic, denoted by $\SliceSetLegacy$, consists of slice flows carried over from a previous time interval and already provisioned on a set of DU-to-CU-UP pairs.
When provisioning new traffic slices, the optimizer may also migrate selected legacy slice flows if doing so reduces overall energy consumption.
Migration can consolidate or redistribute slice flows across servers, thereby reducing active-server energy when its overhead is justified.
Eligibility and delay constraints restrict which DU and CU-UP instances can process each slice.


\subsection{Parameters and Variables}

Let $\ServerSet$ denote the set of servers (indexed by $\ServerIdx$) in the edge cloud.
Each server $\ServerIdx$ has finite resource capacities, specifically
CPU cores ($\Cap_\ServerIdx^{\CPU}$),
memory ($\Cap_\ServerIdx^{\RAM}$),
and NIC I/O bandwidth ($\Cap_\ServerIdx^{\IO}$).
The internal links are provisioned to support the servers' NIC capacities.

\begin{figure}[pos=ht]
  \centering
  \includegraphics[width=0.9\linewidth]{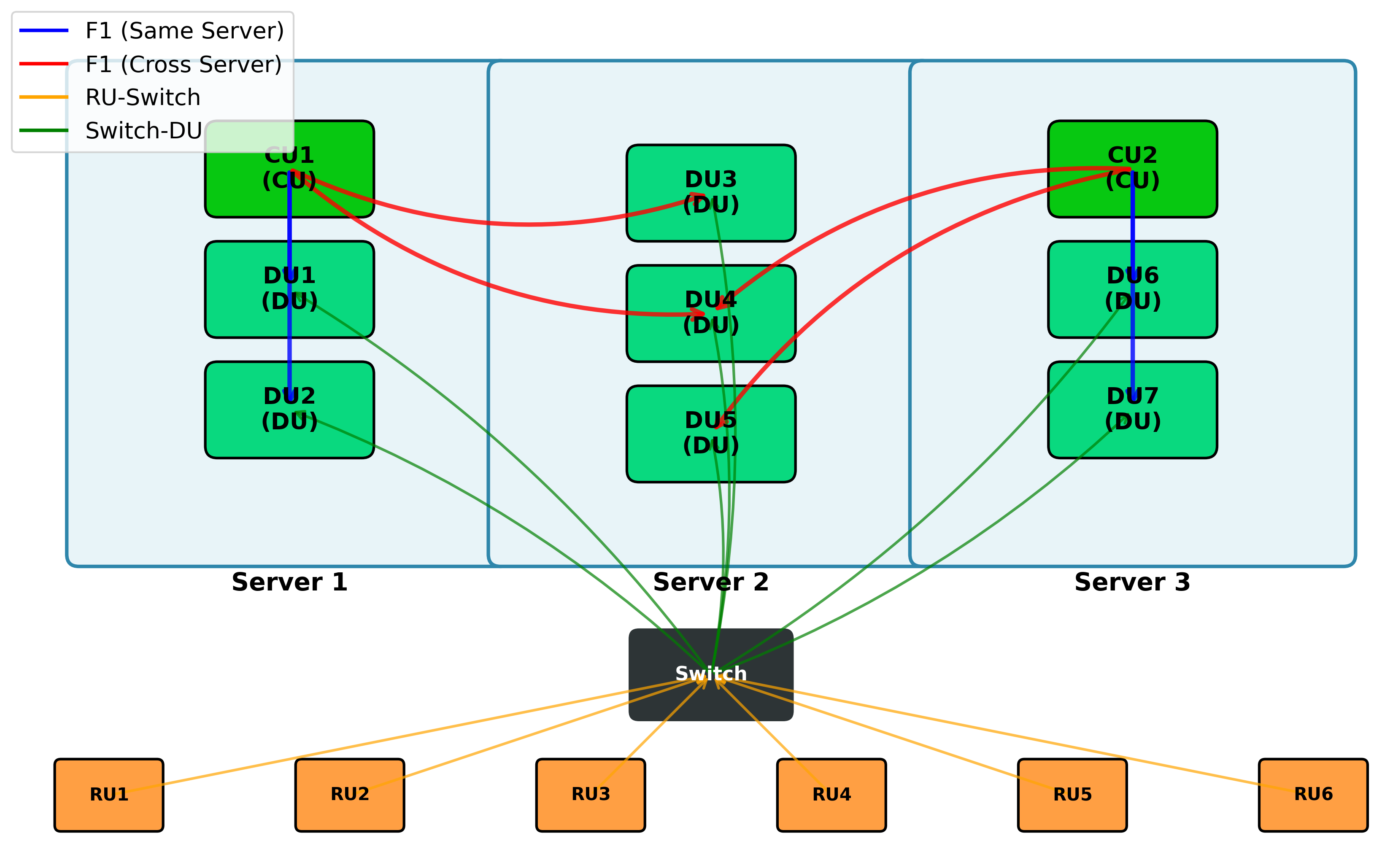}
  \caption{Scenario 2: slice-aware assignment from one DU to multiple possible CU-UP targets under one fixed CU-CP.}
  \label{fig:many_to_many_du_cu_mapping}
\end{figure}

The workload consists of RU-side requests that have reached the edge-cloud ingress and are routed through the internal switching fabric to eligible DU and CU-UP instances.
The switch abstraction in Figs.~\ref{fig:one_to_many_du_cu_mapping} and~\ref{fig:many_to_many_du_cu_mapping} therefore represents only the internal data-center fabric.
To represent heterogeneous slice requirements, we decompose each aggregate traffic-flow request into slice-specific requests at the DU level.
This abstraction is consistent with slice-aware NG-RAN formulations that model distinct throughput, latency, functional-split, placement, and routing requirements for individual network slices~\cite{mushtaq2023optimal,sen2025slice}.
Each request $\FlowIdx \in \FlowSet$ is characterized by
a compute resource vector
$r_{\FlowIdx} = (
  \ResourceIdx_\FlowIdx^{\CPU},
  \ResourceIdx_\FlowIdx^\RAM,
\ResourceIdx_\FlowIdx^\IO)$
which defines the resources needed to provision the corresponding DU and CU-UP functions.
An external preprocessing stage determines the number and sizes of DU/CU-UP CNF instances from predicted demand, consistent with predictive VNF autoscaling approaches~\cite{tran2025proactive,Bo_2024,subramanya2021predictive,verma2024vnfscaling}.
The resulting instance sets and resource vectors are fixed inputs; scaling is not represented by decision variables, constraints, or objective terms in the MILP or heuristic.

We next describe two DU/CU-UP interconnection scenarios: (i) a baseline aligned with current standard capabilities and (ii) an anticipated setting enabled by more advanced traffic-steering and splitting mechanisms. In the current Third Generation Partnership Project (3GPP) framework, user traffic can be differentiated by slice identifiers and associated policies, e.g., Network Slice Selection Assistance Information (NSSAI), Single NSSAI (S-NSSAI), and slice/service profiles.


\subsection{Scenario 1: Each DU connects to a single CU-UP (Single-CU)}

\begin{figure}[pos=ht]
  \centering
  \includegraphics[width=0.9\linewidth]{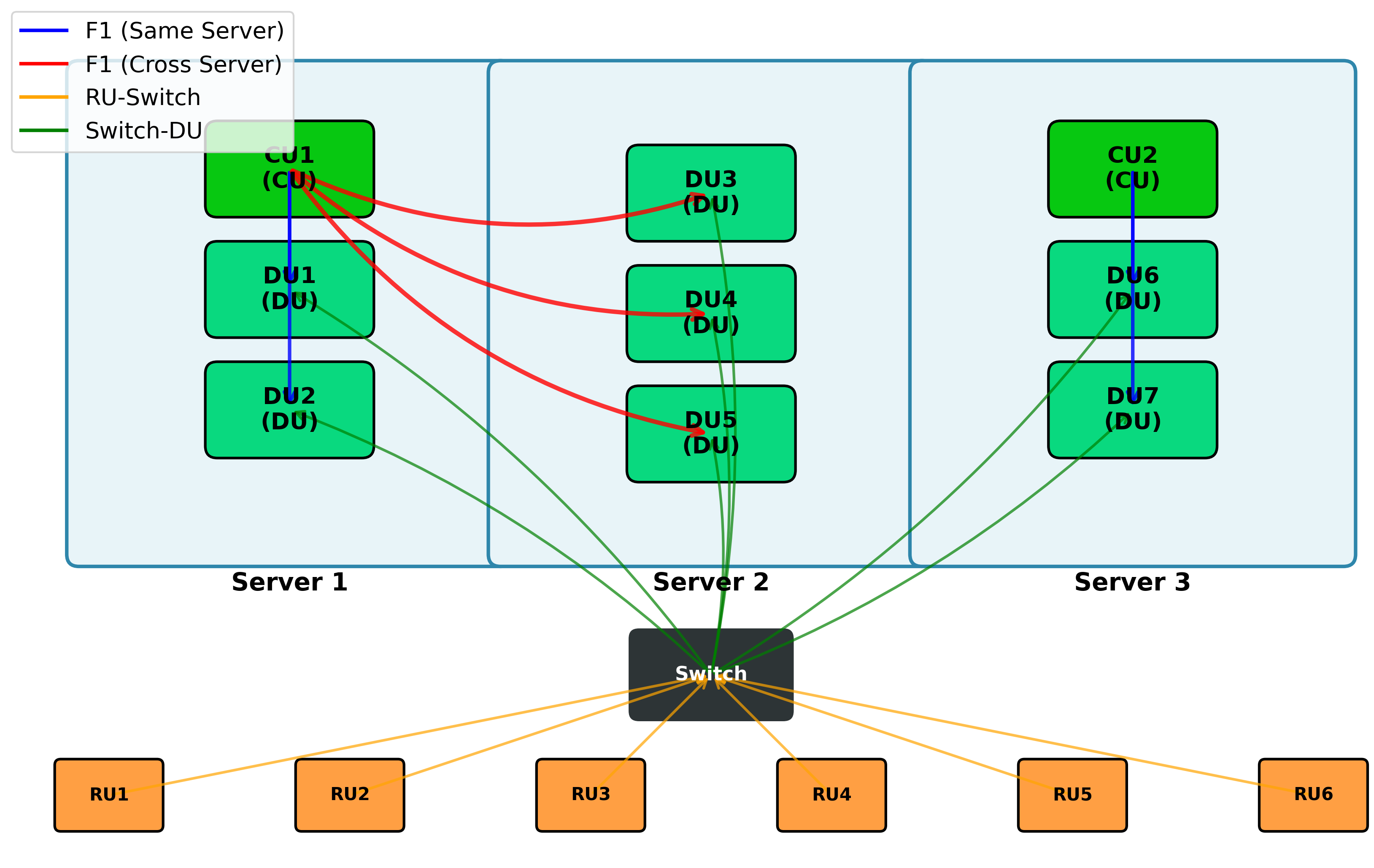}
  \caption{Scenario 1: each DU connects to one CU-UP while retaining one fixed CU-CP association.}
  \label{fig:one_to_many_du_cu_mapping}
\end{figure}
In this scenario, each DU instance uses exactly one CU-UP processing instance, whereas a single CU-UP can concurrently serve multiple DUs (see Fig.~\ref{fig:one_to_many_du_cu_mapping}).
Its fixed CU-CP association over F1-C is not a placement or migration decision.
Each DU $(\DUIdx \in \DUSet)$ and its selected CU-UP $(\CUIdx \in \CUSet)$ may be placed on the same or different servers, provided that resource and slice-specific F1-U delay requirements are satisfied.


\subsection{Scenario 2: Slice-aware DU-to-CU-UP connectivity (Multi-CU)}
\label{sec:scenario_2}

This scenario uses the standardized architectural possibility that one DU can connect to multiple CU-UPs controlled by the same CU-CP~\cite{3gpp38401}.
The mathematical relaxation lies in allowing the optimizer to steer distinct slice-flow groups of that DU to different CU-UP processing targets according to their delay and resource requirements.
Multi-CU therefore means multiple CU-UP targets only: the DU retains one logical CU-CP association, and the model does not place, migrate, or replicate the CU-CP.
Each slice-flow group is assigned to exactly one CU-UP target; therefore, the same traffic is not duplicated, replicated, or broadcast to every CU-UP connected to the DU.
Conversely, each CU-UP target can process slice-flow groups from multiple DUs.
Connectivity in this scenario is defined by a set of slice-specific DU-to-CU-UP pairings ($\DUIdx, \CUIdx$).
For every active slice-specific connection, the joint placement and migration algorithm must ensure that the assigned locations of DU $\DUIdx$ and CU-UP $\CUIdx$ satisfy both delay and resource constraints while reducing energy consumption.

Compared with Scenario 1, this scenario expands the feasible set by allowing the slice-flow groups of one DU to use CU-UP targets on different servers or data-center pods.
The larger feasible set may lower total energy by enabling better server consolidation and load distribution, but it can also increase transmission distance; therefore, the net effect is determined by the common-unit objective rather than assumed in advance.


\section{Minimizing the Overall Energy: MILP Model}
\label{sec:MILP_Formulation}

We formulate the Energy-Aware Joint Placement and Migration (EJPM) problem as a mixed-integer linear program (MILP).
The model is solved once per control interval $t$ using the traffic demand observed for that interval.
The corresponding sets, parameters, decision variables, objective function, and constraints are defined below.


\subsection{Notations}
The sets, parameters, and decision variables used in the formulation are summarized as follows.

\notationheading{Sets and Indices}
\notationrow{$\ServerSet,\DUSet,\CUSet$}{Sets of servers, DUs, and candidate CU-UP processing instances; CU-CP is fixed and omitted.}
\notationrow{$\FlowSet$}{Set of traffic-flow requests, indexed by $\FlowIdx$.}
\notationrow{$\FlowSet_\DUIdx$}{Set of flows pre-assigned to DU $\DUIdx$, with $\bigcup_{\DUIdx}\FlowSet_\DUIdx=\FlowSet$.}
\notationrow{$\SliceSet_\FlowIdx$}{Set of slice-flow groups belonging to flow $\FlowIdx$.}
\notationrow{$\ResourceSet$}{Set of resource types ($\CPU$, $\RAM$, and $\IO$), indexed by $\ResourceIdx$.}
\notationrow{$\PiwiseSet$}{Set of load-dependent incremental-power zones.}

\notationheading{Parameters}

\notationrow{$\DUIdx(\FlowIdx)$}{Pre-assigned DU for flow $\FlowIdx$.}
\notationrow{$\Cap_{\ServerIdx}^{\ResourceIdx}$}{Capacity of resource $\ResourceIdx$ on server $\ServerIdx$.}
\notationrow{$r_{\FlowIdx\ResourceIdx}^{\DU},r_{\FlowIdx\SliceIdx\ResourceIdx}^{\CU}$}{DU-side and CU-UP-side resource requirements.}
\notationrow{$\TrafficFlowData_{\FlowIdx\SliceIdx}$}{Data volume of slice-flow group $(\FlowIdx,\SliceIdx)$ during one control interval (GB).}
\notationrow{$\delta_{\FlowIdx\SliceIdx}^{\mathrm{crit}}$}{One-way DU-to-CU-UP/F1-U delay limit of slice-flow group $(\FlowIdx,\SliceIdx)$.}
\notationrow{$\LatencyBetweenServers$}{One-way network delay between servers $\ServerIdx$ and $\ServerIdx'$.}
\notationrow{$\chi_{\SliceIdx\CUIdx}$}{CU-UP eligibility parameter; 1 if CU-UP $\CUIdx$ can process slice $\SliceIdx$.}
\notationrow{$\MaxCUPerDU$}{Maximum number of distinct CU-UP targets allowed for one DU; the CU-CP count remains one.}
\notationrow{$\ControlInterval$}{Duration of one optimization interval in hours.}
\notationrow{$P_{\ServerIdx}^{\idle},P_{\ServerIdx}^{\mathrm{inc,max}}$}{Fixed idle power and maximum incremental power of server $\ServerIdx$ (W).}
\notationrow{$E_{\ServerIdx}^{\mathrm{wake}}$}{Fixed energy charged when server $\ServerIdx$ changes from inactive to active (kWh).}
\notationrow{$\EnergyLink$}{Transmission-energy coefficient between servers (kWh/GB).}
\notationrow{$B_j^{\mathrm{mig}}$}{Effective state volume transferred when instance $j$ migrates, including retransmitted dirty pages (GB).}
\notationrow{$\beta_{\ServerIdx\ServerIdx'}^{\mathrm{mig}}$}{Effective migration bandwidth from $\ServerIdx$ to $\ServerIdx'$ (GB/s).}
\notationrow{$P_{\ServerIdx}^{\mathrm{src}},P_{\ServerIdx'}^{\mathrm{dst}}$}{Source and destination migration-overhead power (W).}
\notationrow{$\epsilon_{\ServerIdx\ServerIdx'}^{\mathrm{net}}$}{Network energy per transferred GB on the migration path (kWh/GB).}
\notationrow{$\MigrationCoeff_{j\ServerIdx\ServerIdx'}$}{Precomputed migration energy of instance $j$ from $\ServerIdx$ to $\ServerIdx'$ (kWh).}
\notationrow{$\PreviousPlacementDU,\PreviousPlacementCU$}{Previous-interval DU and CU-UP placement parameters.}
\notationrow{$\PreviousActive_{\ServerIdx}$}{Previous-interval server-active parameter.}
\notationrow{$w_{\ResourceIdx}$}{Weight of resource $\ResourceIdx$ in aggregate server utilization.}
\notationrow{$\underline{\Utilization}_{\PiwiseIdx},\overline{\Utilization}_{\PiwiseIdx}$}{Utilization bounds of incremental-power zone $\PiwiseIdx$.}
\notationrow{$a_{\PiwiseIdx}(\Temperature),b_{\PiwiseIdx}(\Temperature)$}{Incremental-power slope and intercept in zone $\PiwiseIdx$ (W).}
\notationrow{$M_{\ServerIdx\PiwiseIdx}^{P}$}{Valid big-$M$ bound for the incremental-power equality in zone $\PiwiseIdx$.}


\notationheading{Decision Variables}

\notationrow{$x^\textsc{DU}_{\DUIdx \ServerIdx},$\\$x^\textsc{CU}_{\CUIdx \ServerIdx}$}{Binary placement variables; equal to 1 if DU $\DUIdx$ (resp. CU-UP $\CUIdx$) is placed on server $\ServerIdx$.}
\notationrow{$\AssignCU_{\FlowIdx\SliceIdx\CUIdx}$}{Binary assignment variable; equal to 1 if slice-flow group $(\FlowIdx,\SliceIdx)$ is processed by CU-UP $\CUIdx$.}
\notationrow{$y_{\FlowIdx\SliceIdx\CUIdx\ServerIdx\ServerIdx'}$}{Binary route variable; equal to 1 if $(\FlowIdx,\SliceIdx)$ is transported from its DU on $\ServerIdx$ to CU-UP $\CUIdx$ on $\ServerIdx'$.}
\notationrow{$z_{\DUIdx\CUIdx}$}{Binary logical-connection variable; equal to 1 if there is an active user-plane connection between DU $\DUIdx$ and CU-UP $\CUIdx$.}
\notationrow{$\MigrationIdx_{\DUIdx \ServerIdx \ServerIdx'}^\textsc{DU},$\\$\MigrationIdx_{\CUIdx \ServerIdx \ServerIdx'}^\textsc{CU}$}{Binary migration indicators; equal to 1 if DU $\DUIdx$ (resp. CU-UP $\CUIdx$) migrates from server $\ServerIdx$ to $\ServerIdx'$.}
\notationrow{$\VarActive_{\ServerIdx}$}{Binary server-activation variable; equal to 1 if server $\ServerIdx$ is active.}
\notationrow{$\WakeUp_{\ServerIdx}$}{Binary wake-up variable; equal to 1 only when server $\ServerIdx$ changes from inactive to active.}
\notationrow{$\Utilization_{\ServerIdx}$}{Nonnegative continuous variable representing the weighted aggregate utilization of server $\ServerIdx$.}
\notationrow{$\PiwisePower_{\ServerIdx}$}{Nonnegative continuous variable representing load-dependent incremental server power (W).}
\notationrow{$\gamma_{\ServerIdx\PiwiseIdx}$}{Binary zone-selection variable; equal to 1 if server $\ServerIdx$ operates in zone $\PiwiseIdx$.}

\subsection{Objective Function}
All objective terms are expressed in kWh. The MILP minimizes total interval energy without an arbitrary weighting parameter:

\begin{equation}
  \label{eq:total_objective}
  \min \quad E^{\mathrm{total}} = E^{\Placement}+E^{\Migration}.
\end{equation}

The operational term $E^{\Placement}$ includes fixed idle power, load-dependent incremental power, fixed wake-up energy, and traffic-transmission energy:

\begin{multline}
  \label{eq:energy_placement}
  E^{\Placement} =
  \frac{\ControlInterval}{1000}
  \sum_{\ServerIdx \in \ServerSet}
  \left(P_{\ServerIdx}^{\idle}\VarActive_{\ServerIdx}+\PiwisePower_{\ServerIdx}\right)
  +\sum_{\ServerIdx \in \ServerSet}E_{\ServerIdx}^{\mathrm{wake}}\WakeUp_{\ServerIdx}\\
  +\sum_{\FlowIdx \in \FlowSet}\sum_{\SliceIdx \in \SliceSet_{\FlowIdx}}
  \sum_{\CUIdx \in \CUSet}\sum_{\ServerIdx,\ServerIdx' \in \ServerSet}
  \TrafficFlowData_{\FlowIdx\SliceIdx}\EnergyLink
  y_{\FlowIdx\SliceIdx\CUIdx\ServerIdx\ServerIdx'}.
\end{multline}

Here, $\ControlInterval$ is measured in hours, server power is measured in watts, and division by 1000 converts Wh to kWh. The wake-up term is a fixed input cost incurred once for each off-to-on transition; it is not multiplied by the interval duration.

\noindent\textit{Load-dependent incremental server power}
\label{sec:piecewise_power}
We separate fixed idle power from the load-dependent incremental component. Thus, $\PiwisePower_{\ServerIdx}$ is zero when server $\ServerIdx$ is inactive and contains no idle/base-power contribution.
For each $\ServerIdx\in\ServerSet$, the weighted utilization is

\begin{multline}
  \Utilization_{\ServerIdx} =
  \sum_{\ResourceIdx \in \ResourceSet}w_{\ResourceIdx}
  \frac{
    \sum_{\FlowIdx \in \FlowSet}
    r_{\FlowIdx\ResourceIdx}^{\DU}x^{\DU}_{\DUIdx(\FlowIdx)\ServerIdx}
  }{\Cap_{\ServerIdx}^{\ResourceIdx}}\\
  +\sum_{\ResourceIdx \in \ResourceSet}w_{\ResourceIdx}
  \frac{
    \sum_{\FlowIdx \in \FlowSet}\sum_{\SliceIdx \in \SliceSet_{\FlowIdx}}
    \sum_{\CUIdx \in \CUSet}\sum_{\ServerIdx' \in \ServerSet}
    r_{\FlowIdx\SliceIdx\ResourceIdx}^{\CU}
    y_{\FlowIdx\SliceIdx\CUIdx\ServerIdx'\ServerIdx}
  }{\Cap_{\ServerIdx}^{\ResourceIdx}}.
\end{multline}

The following explicit inactive-server bounds force utilization and incremental power to zero whenever $\VarActive_{\ServerIdx}=0$:
\begin{align}
  &0\le \Utilization_{\ServerIdx}\le \MaxUtil\VarActive_{\ServerIdx}
  &&\forall \ServerIdx\in\ServerSet,\label{eq:inactive_util_bound}\\
  &0\le \PiwisePower_{\ServerIdx}\le
  P_{\ServerIdx}^{\mathrm{inc,max}}\VarActive_{\ServerIdx}
  &&\forall \ServerIdx\in\ServerSet.\label{eq:inactive_power_bound}
\end{align}

Exactly one incremental-power zone is selected for an active server:
\begin{align}
  &\sum_{\PiwiseIdx\in\PiwiseSet}\gamma_{\ServerIdx\PiwiseIdx}
  =\VarActive_{\ServerIdx}
  &&\forall \ServerIdx\in\ServerSet,\label{eq:one_zone_active}\\
  &\Utilization_{\ServerIdx}\ge
  \underline{\Utilization}_{\PiwiseIdx}\gamma_{\ServerIdx\PiwiseIdx}
  &&\forall \ServerIdx,\PiwiseIdx,\label{eq:utilization_zone_lb}\\
  &\Utilization_{\ServerIdx}\le
  \overline{\Utilization}_{\PiwiseIdx}\gamma_{\ServerIdx\PiwiseIdx}
  +\MaxUtil(1-\gamma_{\ServerIdx\PiwiseIdx})
  &&\forall \ServerIdx,\PiwiseIdx.\label{eq:utilization_zone_ub}
\end{align}

For the selected zone, the incremental power is equal to its affine expression. Both inequalities are required; objective minimization alone is not used as a substitute for the upper equality:
\begin{align}
  \PiwisePower_{\ServerIdx}
  &\ge a_{\PiwiseIdx}(\Temperature)\Utilization_{\ServerIdx}
  +b_{\PiwiseIdx}(\Temperature)
  -M_{\ServerIdx\PiwiseIdx}^{P}(1-\gamma_{\ServerIdx\PiwiseIdx}),
  \label{eq:power_zone_lb}\\
  \PiwisePower_{\ServerIdx}
  &\le a_{\PiwiseIdx}(\Temperature)\Utilization_{\ServerIdx}
  +b_{\PiwiseIdx}(\Temperature)
  +M_{\ServerIdx\PiwiseIdx}^{P}(1-\gamma_{\ServerIdx\PiwiseIdx})
  \label{eq:power_max_of_zones}
\end{align}
for all $\ServerIdx\in\ServerSet$ and $\PiwiseIdx\in\PiwiseSet$.
Temperature is observed before each control interval and is therefore a fixed input to the MILP; no product of decision variables is introduced.

\noindent\textit{Physically calibrated migration energy}
For any DU or CU instance $j$ migrating from $\ServerIdx$ to $\ServerIdx'$, the effective migration duration and energy coefficient are precomputed as
\begin{align}
  \tau_{j\ServerIdx\ServerIdx'}^{\mathrm{mig}}
  &=\frac{B_j^{\mathrm{mig}}}{\beta_{\ServerIdx\ServerIdx'}^{\mathrm{mig}}},
  \label{eq:migration_duration}\\
  \MigrationCoeff_{j\ServerIdx\ServerIdx'}
  &=B_j^{\mathrm{mig}}\epsilon_{\ServerIdx\ServerIdx'}^{\mathrm{net}}
  +\frac{\left(P_{\ServerIdx}^{\mathrm{src}}+P_{\ServerIdx'}^{\mathrm{dst}}\right)
  \tau_{j\ServerIdx\ServerIdx'}^{\mathrm{mig}}}{3.6\times10^{6}}.
  \label{eq:migration_coefficient}
\end{align}
$B_j^{\mathrm{mig}}$ includes retransmitted dirty pages, $\beta^{\mathrm{mig}}$ is in GB/s, and the second term converts watt-seconds to kWh. Because every $\MigrationCoeff_{j\ServerIdx\ServerIdx'}$ is computed from observed or calibrated inputs before optimization, the migration objective remains linear:
\begin{multline}
  \label{eq:migration_cost}
  E^{\Migration}=
  \sum_{\DUIdx\in\DUSet}\sum_{\substack{\ServerIdx,\ServerIdx'\in\ServerSet\\\ServerIdx\neq\ServerIdx'}}
  \MigrationCoeff_{\DUIdx\ServerIdx\ServerIdx'}^{\DU}
  \MigrationIdx_{\DUIdx\ServerIdx\ServerIdx'}^{\DU}\\
  +\sum_{\CUIdx\in\CUSet}\sum_{\substack{\ServerIdx,\ServerIdx'\in\ServerSet\\\ServerIdx\neq\ServerIdx'}}
  \MigrationCoeff_{\CUIdx\ServerIdx\ServerIdx'}^{\CU}
  \MigrationIdx_{\CUIdx\ServerIdx\ServerIdx'}^{\CU}.
\end{multline}

To ensure migration variables cannot take value $1$ unless a move occurs, the following inequalities hold for every instance and every $\ServerIdx,\ServerIdx'\in\ServerSet$ with $\ServerIdx\neq\ServerIdx'$:

\begin{align}
  & \MigrationIdx_{\DUIdx\ServerIdx\ServerIdx'}^{\DU} \le \PreviousPlacement_{\DUIdx\ServerIdx}^{\DU},
  && \forall \DUIdx \in \DUSet,\label{eq:migr_du_prev}\\
  & \MigrationIdx_{\DUIdx\ServerIdx\ServerIdx'}^{\DU} \le x^{\DU}_{\DUIdx\ServerIdx'},
  && \forall \DUIdx \in \DUSet,\label{eq:migr_du_tight}\\
  & \MigrationIdx_{\CUIdx\ServerIdx\ServerIdx'}^{\CU} \le \PreviousPlacement_{\CUIdx\ServerIdx}^{\CU},
  && \forall \CUIdx \in \CUSet,\label{eq:migr_cu_prev}\\
  & \MigrationIdx_{\CUIdx\ServerIdx\ServerIdx'}^{\CU} \le x^{\CU}_{\CUIdx\ServerIdx'},
  && \forall \CUIdx \in \CUSet.\label{eq:migr_cu_tight}
\end{align}

\subsection{Constraints}

To strengthen the formulation and explicitly link the main decision variables, we include the following standard constraints.
\medskip

\noindent\textit{Placement, activation, and wake-up constraints.}
A server is active exactly when it hosts at least one DU or CU:

\begin{align}
  & \VarActive_{\ServerIdx} \ge x^{\DU}_{\DUIdx\ServerIdx}
  && \forall \DUIdx, \ServerIdx \label{eq:active_link_du}\\
  & \VarActive_{\ServerIdx} \ge x^{\CU}_{\CUIdx\ServerIdx}
  && \forall \CUIdx,\ServerIdx \label{eq:active_link_cu}\\
  & \VarActive_\ServerIdx \le \sum_\DUIdx x^{\DU}_{\DUIdx\ServerIdx}+\sum_\CUIdx x^{\CU}_{\CUIdx\ServerIdx} \label{eq:active_server_cu_du}
\end{align}

The binary wake-up variable is one exactly when a previously inactive server becomes active:
\begin{align}
  &\WakeUp_{\ServerIdx}\ge \VarActive_{\ServerIdx}-\PreviousActive_{\ServerIdx}
  &&\forall \ServerIdx\in\ServerSet,\label{eq:wake_lb}\\
  &\WakeUp_{\ServerIdx}\le \VarActive_{\ServerIdx},\qquad
  \WakeUp_{\ServerIdx}\le 1-\PreviousActive_{\ServerIdx}
  &&\forall \ServerIdx\in\ServerSet.\label{eq:wake_ub}
\end{align}

Every DU and CU must be assigned to exactly one server:
\begin{alignat}{2}
  & \sum_{\ServerIdx \in \ServerSet} x^{\DU}_{\DUIdx \ServerIdx} = 1 \quad && \forall \DUIdx \in \DUSet \label{eq:du_placement} \\
  & \sum_{\ServerIdx \in \ServerSet} x^{\CU}_{\CUIdx \ServerIdx} = 1 \quad && \forall \CUIdx \in \CUSet \label{eq:cu_placement}
\end{alignat}

The total DU-side and CU-side resource consumption on any server must not exceed its capacity:
\begin{multline}
  \sum_{\FlowIdx\in\FlowSet}
  r_{\FlowIdx\ResourceIdx}^{\DU}x^{\DU}_{\DUIdx(\FlowIdx)\ServerIdx}
  +\sum_{\FlowIdx\in\FlowSet}\sum_{\SliceIdx\in\SliceSet_{\FlowIdx}}
  \sum_{\CUIdx\in\CUSet}\sum_{\ServerIdx'\in\ServerSet}
  r_{\FlowIdx\SliceIdx\ResourceIdx}^{\CU}
  y_{\FlowIdx\SliceIdx\CUIdx\ServerIdx'\ServerIdx}
  \le \Cap_{\ServerIdx}^{\ResourceIdx}\, \VarActive_{\ServerIdx} \\
  \quad \forall \ServerIdx \in \ServerSet,\ \forall \ResourceIdx \in \ResourceSet.
\end{multline}

\noindent\textit{Slice-flow assignment and placement coupling.}
Every slice-flow group is assigned to exactly one eligible CU:
\begin{align}
  &\sum_{\CUIdx\in\CUSet}\AssignCU_{\FlowIdx\SliceIdx\CUIdx}=1
  &&\forall \FlowIdx\in\FlowSet,\ \SliceIdx\in\SliceSet_{\FlowIdx},
  \label{eq:one_cu_per_slice_flow}\\
  &\AssignCU_{\FlowIdx\SliceIdx\CUIdx}\le \chi_{\SliceIdx\CUIdx}
  &&\forall \FlowIdx,\SliceIdx,\CUIdx.
  \label{eq:cu_eligibility}
\end{align}

The CU-indexed route variable is equal to the selected assignment and is linked to both endpoint placements:
\begin{align}
  &\sum_{\ServerIdx,\ServerIdx'\in\ServerSet}
  y_{\FlowIdx\SliceIdx\CUIdx\ServerIdx\ServerIdx'}
  =\AssignCU_{\FlowIdx\SliceIdx\CUIdx}
  &&\forall \FlowIdx,\SliceIdx,\CUIdx,\label{eq:y_equals_q}\\
  &y_{\FlowIdx\SliceIdx\CUIdx\ServerIdx\ServerIdx'}
  \le x^{\DU}_{\DUIdx(\FlowIdx)\ServerIdx}
  &&\forall \FlowIdx,\SliceIdx,\CUIdx,\ServerIdx,\ServerIdx',
  \label{eq:y_le_xdu}\\
  &y_{\FlowIdx\SliceIdx\CUIdx\ServerIdx\ServerIdx'}
  \le x^{\CU}_{\CUIdx\ServerIdx'}
  &&\forall \FlowIdx,\SliceIdx,\CUIdx,\ServerIdx,\ServerIdx'.
  \label{eq:y_le_xcu}
\end{align}

The logical DU--CU variable is linked bidirectionally to the slice-flow assignments:
\begin{align}
  &z_{\DUIdx(\FlowIdx)\CUIdx}\ge
  \AssignCU_{\FlowIdx\SliceIdx\CUIdx}
  &&\forall \FlowIdx,\SliceIdx,\CUIdx,\label{eq:z_from_q_lb}\\
  &z_{\DUIdx\CUIdx}\le
  \sum_{\FlowIdx\in\FlowSet_{\DUIdx}}\sum_{\SliceIdx\in\SliceSet_{\FlowIdx}}
  \AssignCU_{\FlowIdx\SliceIdx\CUIdx}
  &&\forall \DUIdx,\CUIdx.\label{eq:z_from_q_ub}
\end{align}

Both scenarios use the same formulation. Setting $\MaxCUPerDU=1$ yields Single-CU-UP (Single-CU), whereas $\MaxCUPerDU>1$ yields the Multi-CU-UP relaxation (Multi-CU). In both cases, the DU retains one fixed CU-CP association:

\begin{equation}
  \sum_{\CUIdx \in \CUSet} z_{\DUIdx \CUIdx} \leq \MaxCUPerDU \quad \forall \DUIdx \in \DUSet
\end{equation}

\noindent\textit{Latency Requirements}
For every $\FlowIdx\in\FlowSet$ and $\SliceIdx\in\SliceSet_{\FlowIdx}$, the modelled one-way DU-to-CU-UP/F1-U latency must satisfy

\begin{multline}
  \sum_{\CUIdx\in\CUSet}\sum_{\ServerIdx,\ServerIdx'\in\ServerSet}
  \left(\LatencyBetweenServers+\delta_{\ServerIdx}^{\mathrm{proc,DU}}
  +\delta_{\ServerIdx'}^{\mathrm{proc,CU}}\right)\\
  {}\cdot y_{\FlowIdx\SliceIdx\CUIdx\ServerIdx\ServerIdx'}
  \le \delta_{\FlowIdx\SliceIdx}^{\mathrm{crit}}.
\end{multline}

Here, $\LatencyBetweenServers$ denotes the one-way intra-edge-cloud network delay between the servers hosting the DU and CU-UP instances. The endpoint terms include only DU-to-CU-UP/F1-U processing; they do not include F1-C control signaling, RU-DU fronthaul, CU-to-core transport, or application/core processing latency.

\noindent\textit{Migration Logic}
Migration is detected by comparing the current placement ($x$) with the previous placement parameters ($\PreviousPlacementDU$ and $\PreviousPlacementCU$).
If a DU or CU is placed on a server different from its placement at time $t-1$, the corresponding migration variable is set to 1:

\begin{alignat}{2}
  & \MigrationIdx_{\DUIdx\ServerIdx\ServerIdx'}^{\DU} \geq \PreviousPlacementDU + x^{\DU}_{\DUIdx \ServerIdx'} - 1 \qquad
  && \DUIdx \in \DUSet, \ServerIdx \neq \ServerIdx' \in \ServerSet \\
  & \MigrationIdx_{\CUIdx \ServerIdx\ServerIdx'}^{\CU} \geq \PreviousPlacement_{\CUIdx\ServerIdx}^{\CU} + x_{\CUIdx\ServerIdx'}^{\CU} - 1
  && \CUIdx \in \CUSet, \ServerIdx \neq \ServerIdx' \in \ServerSet
\end{alignat}


\section{$k$-means-Based Heuristic for Joint DU/CU-UP Placement and Migration}

We now present a deterministic energy-aware heuristic ({\NameHeuristic}) for repeated DU/CU-UP placement and migration decisions. It avoids solving the NP-hard placement problem to proven optimality at every interval by decomposing each decision into four phases:

\begin{itemize}
  \item [1.] Slice-flow-to-CU-UP assignment using normalized $k$-means++ and deterministic CU-UP matching.
  \item [2.] Delay-aware physical mapping using ordered server clusters and explicit incremental-energy costs.
  \item [3.] Sequential migration validation and consolidation.
  \item [4.] Post-processing feasibility repair and violation reporting.
\end{itemize}

All randomization is controlled by a fixed seed and a fixed number of $k$-means restarts. The repair phase attempts to remove residual violations, but it does not provide a global completeness guarantee; any remaining violation is reported explicitly.

\subsection{Four-Phase Heuristic Algorithm}

The heuristic uses the same slice-flow assignment, placement, wake-up, transmission, and migration-energy definitions as the MILP in Section~\ref{sec:MILP_Formulation}. Time indices are omitted because the procedure is executed once per control interval using the previous interval as input state.

\subsubsection{Heuristic-specific notation}
Table~\ref{tab:heuristic_notation} defines the additional heuristic-specific symbols and hyperparameters used by the main procedure and its four phase-specific algorithms.

\begin{table}[pos=t]
  \centering
  \caption{Heuristic-specific notation and hyperparameters}
  \label{tab:heuristic_notation}
  \begin{tabularx}{\columnwidth}{>{\raggedright\arraybackslash}m{0.27\columnwidth}>{\raggedright\arraybackslash}X}
    \toprule
    \textbf{Notation} & \textbf{Definition} \\
    \midrule
    $N_{\mathrm{sf}}$ & Number of active slice-flow groups, $N_{\mathrm{sf}}=\sum_{\FlowIdx\in\FlowSet}|\SliceSet_{\FlowIdx}|$. \\
    $\eta_{\CU/\DU}$ & Slice-flow grouping ratio used to derive the Phase-1 cluster count. \\
    $R_k,I_k$ & Fixed number of $k$-means restarts and maximum iterations per restart. \\
    $\xi$ & Fixed pseudo-random seed used by every clustering call. \\
    $I_{\mathrm{rep}}$ & Maximum number of repair iterations in Phase 4; integer hyperparameter ($I_{\mathrm{rep}} \in \mathbb{Z}_{\ge 0}$). \\
    $|\mathrm{pods}|$ & Number of server clusters used in Phase 2. \\
    $\epsilon$ & Minimum strict energy improvement required to accept a move (kWh). \\
    $u_{\mathrm{cons}}$ & Server-utilization threshold below which a server is considered a consolidation candidate in Phase 3. \\
    $\Delta E(j,s')$ & Total-energy change in kWh when instance $j$ is moved to server $s'$. \\
    $\nu$ & Phase-4 violation report tuple $(r_{\mathrm{pre}}, r_{\mathrm{post}}, \bar{v}_{\mathrm{post}}, v_{\mathrm{post}}^{\max}, n_{\mathrm{rep}})$. \\
    \bottomrule
  \end{tabularx}
\end{table}

\begin{varalgorithm}{\NameHeuristic}
  \caption{Joint Placement and Migration Heuristic}
  \label{alg:main_four_phases}

  \begin{algorithmic}[1]
    \Statex \textbf{Input:} $\ServerSet$, $\FlowSet$, $\CUSet$, delay matrix $\delta$, previous placement $\PreviousPlacement$,
    \Statex \hspace*{1.cm} migration coefficients $\MigrationCoeff$, wake-up costs $E^{\mathrm{wake}}$, $R_k$, $I_k$, seed $\xi$, ratio $\eta_{\CU/\DU}$,
    \Statex \hspace*{1.cm} limit $\MaxCUPerDU$, pod count $|\mathrm{pods}|$, $I_{\mathrm{rep}}$, $\epsilon$, $u_{\mathrm{cons}}$
    \Statex \textbf{Output:} Placement $x$, assignment $q$, pairings $z$, migrations $m$, report $\nu$
    \Statex
    \State Build slice-flow set $\mathcal{F}\gets\{(\FlowIdx,\SliceIdx):\FlowIdx\in\FlowSet,\SliceIdx\in\SliceSet_{\FlowIdx}\}$
    \State $k_{\mathrm{pair}}\gets\min\{|\CUSet|,\max\{1,\lceil|\mathcal{F}|/\eta_{\CU/\DU}\rceil\}\}$
    \Statex
    \State \textit{// Phase 1: Logical Pairing}
    \State $q,z \gets \text{PairFlowCUUP}(\mathcal{F},\CUSet,k_{\mathrm{pair}},R_k,I_k,\xi,\MaxCUPerDU)$
    \Statex
    \State \textit{// Phase 2: Delay-Aware Physical Mapping}
    \State $x^*,\mathcal{U} \gets \text{MapToPhysical}(q,z,\ServerSet,\delta,|\mathrm{pods}|,\PreviousPlacement,\MigrationCoeff,E^{\mathrm{wake}})$
    \Statex
    \State \textit{// Phase 3: Energy-Efficient Migration \& Consolidation}
    \State $x,m \gets \text{MigrateSequential}(x^*,\PreviousPlacement,q,z,\MigrationCoeff,E^{\mathrm{wake}},\epsilon,u_{\mathrm{cons}})$
    \Statex
    \State \textit{// Phase 4: Post-Processing Feasibility Repair}
    \State $x,q,z,m,\nu \gets \text{RepairFeasibility}(x,q,z,m,\mathcal{U},I_{\mathrm{rep}})$
    \Statex \Return $(x,q,z,m,\nu)$
  \end{algorithmic}
\end{varalgorithm}


%
\begin{algorithm}[h]
  \small
  \caption{Phase 1: PairFlowCU-UP (Deterministic Slice-Flow Assignment)}
  \label{alg:phase1_kmeans_du_cu}

  \begin{algorithmic}[1]
    \Statex \textbf{Input:} Slice-flow set $\mathcal{F}$, CU-UP pool $\CUSet$, $k_{\mathrm{pair}}$, $R_k$, $I_k$, seed $\xi$, limit $\MaxCUPerDU$
    \Statex \textbf{Output:} Assignment $q$ and logical pairing matrix $z$
    \Statex
    \For{each $(\FlowIdx,\SliceIdx)\in\mathcal{F}$}
    \State $f_{\FlowIdx\SliceIdx}\gets[\delta_{\FlowIdx\SliceIdx}^{\mathrm{crit}},
      r_{\FlowIdx\SliceIdx\CPU}^{\CU},
      r_{\FlowIdx\SliceIdx\RAM}^{\CU},
    r_{\FlowIdx\SliceIdx\IO}^{\CU}]$
    \EndFor
    \State Min--max normalize each feature; set a constant feature to zero
    \For{$r=0$ to $R_k-1$}
    \State Run $k$-means++ with $k_{\mathrm{pair}}$ clusters, seed $\xi+r$, and at most $I_k$ iterations
    \EndFor
    \State Keep the clustering with minimum within-cluster sum of squares
    \State Build $H_{\ell\CUIdx}$ from normalized capacity mismatch and eligibility penalties
    \State Match clusters to distinct CU-UPs using the Hungarian algorithm
    \State Initialize $q\gets 0$
    \Statex
    \If{$\MaxCUPerDU=1$}
    \For{each DU $\DUIdx$}
    \State Assign all $(\FlowIdx,\SliceIdx)$ with $\FlowIdx\in\FlowSet_{\DUIdx}$ to the eligible CU-UP of minimum aggregate $H$ cost
    \EndFor
    \Else
    \State Assign each slice-flow group to its matched eligible CU-UP
    \For{each DU using more than $\MaxCUPerDU$ distinct CU-UPs}
    \State Keep the $\MaxCUPerDU$ CU-UPs of minimum aggregate cost and reassign the remaining groups to their lowest-cost retained CU-UP
    \EndFor
    \EndIf
    \State Set $z_{\DUIdx\CUIdx}=1$ iff some assigned group of DU $\DUIdx$ uses CU-UP $\CUIdx$
    \Statex \textbf{Return} $(q,z)$
  \end{algorithmic}
\end{algorithm}
%

\begin{figure*}[pos=t]
  \centering
  \includegraphics[width=0.98\textwidth]{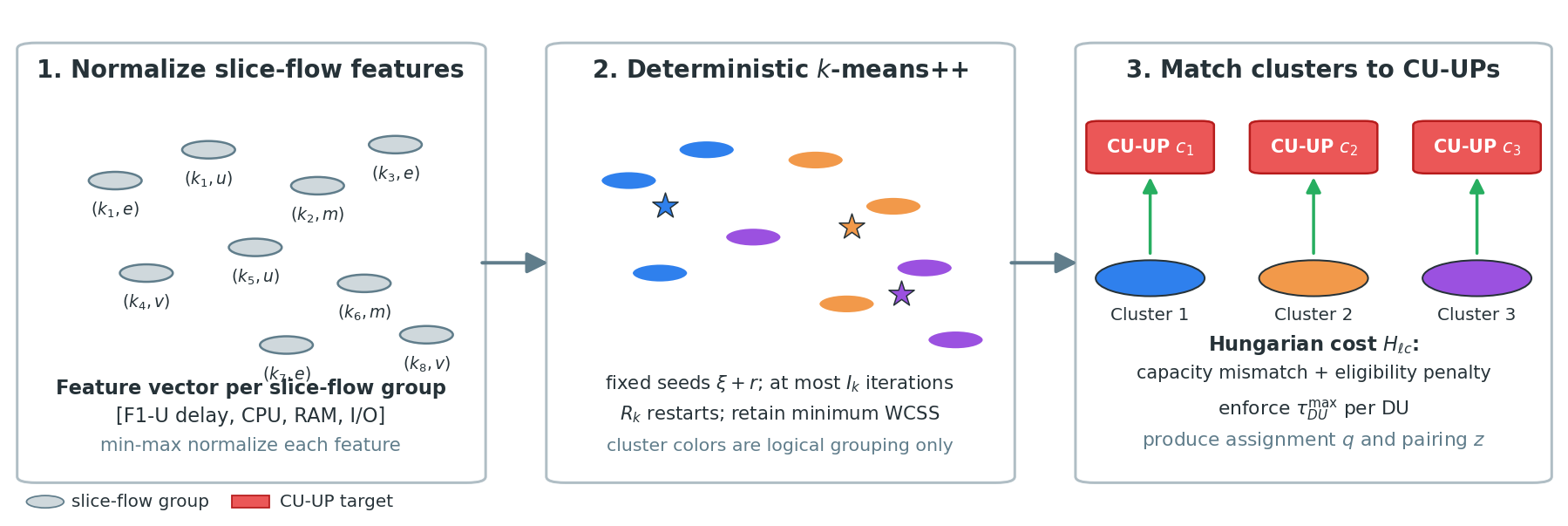}
  \caption{Illustration of the Phase-1 logical pairing procedure. Slice-flow groups are represented by normalized F1-U-delay and CU-UP resource-demand features, deterministically clustered, and matched to eligible CU-UPs using the Hungarian algorithm. The resulting assignments are adjusted, when necessary, to satisfy the maximum number of CU-UP associations per DU.}
  \label{fig:k_means_clustering}
\end{figure*}

In Phase 1, Algorithm~\ref{alg:phase1_kmeans_du_cu} clusters slice-flow groups rather than whole DUs, because the assignment variable $\AssignCU_{\FlowIdx\SliceIdx\CUIdx}$ is slice specific. Fixed restarts, iteration limits, and seeds make the output reproducible.
As illustrated in Fig.~\ref{fig:k_means_clustering}, clustering provides a reproducible grouping aid; it does not replace eligibility, connectivity, or capacity checks. Each slice-flow feature combines its delay limit and CU-UP-side CPU, RAM, and I/O demands. For each feature $x$, min--max normalization is

\begin{equation}
  x' =
  \begin{cases}
    \dfrac{x-\min(x)}{\max(x)-\min(x)}, & \max(x)>\min(x),\\
    0, & \max(x)=\min(x).
  \end{cases}
\end{equation}

\noindent
The Hungarian cost $H_{\ell\CUIdx}$ combines normalized CU-UP-capacity mismatch with a prohibitive penalty for an ineligible slice--CU-UP pair. Single-CU and Multi-CU use the same assignment procedure; only $\MaxCUPerDU$ changes. No reliability requirement is introduced unless it is explicitly present in the optimization inputs.

%
\begin{algorithm}[h]
  \small
  \caption{Phase 2: MapToPhysical (Deterministic Delay-Aware Placement)}
  \label{alg:map_to_physical_servers}

  \begin{algorithmic}[1]
    \Statex \textbf{Input:} Assignment $q$, pairing $z$, servers $\ServerSet$, delay matrix $\delta$, cluster count $k_s$, previous placement $\PreviousPlacement$,
    \Statex \hspace*{1.cm} migration coefficients $\MigrationCoeff$, wake-up costs $E^{\mathrm{wake}}$, and the MILP server/transmission-energy parameters
    \Statex \textbf{Output:} Candidate placement $x^*$ and unassigned-instance set $\mathcal{U}$
    \Statex
    \State Cluster servers using fixed features $[\mathrm{pod}(\ServerIdx),\mathrm{mean}_{\ServerIdx'}\delta_{\ServerIdx\ServerIdx'},\max_{\ServerIdx'}\delta_{\ServerIdx\ServerIdx'}]$
    \State Order $j\in\DUSet\cup\CUSet$ by decreasing normalized resource demand, then increasing strictest incident delay, then instance ID
    \State Initialize $x^*\gets0$, residual capacities, and $\mathcal{U}\gets\emptyset$
    \For{each ordered instance $j$}
    \State Visit server clusters in increasing distance from the previous cluster of $j$
    \State Build candidate set $\mathcal{S}_j$ satisfying residual capacity and every currently determined DU-to-CU-UP delay constraint
    \For{each $\ServerIdx\in\mathcal{S}_j$}
    \State Compute $C(j,\ServerIdx)\gets\Delta E^{\mathrm{server}}+\Delta E^{\mathrm{tx}}+\Delta E^{\mathrm{wake}}+\Delta E^{\mathrm{mig}}$
    \EndFor
    \If{$\mathcal{S}_j\neq\emptyset$}
    \State Place $j$ on the minimum-cost server; break ties by server ID; update residual capacity
    \Else
    \State $\mathcal{U}\gets\mathcal{U}\cup\{j\}$
    \EndIf
    \EndFor
    \Statex \textbf{Return} $(x^*,\mathcal{U})$
  \end{algorithmic}
\end{algorithm}

\begin{figure*}[pos=t]
  \centering
  \includegraphics[width=0.98\textwidth]{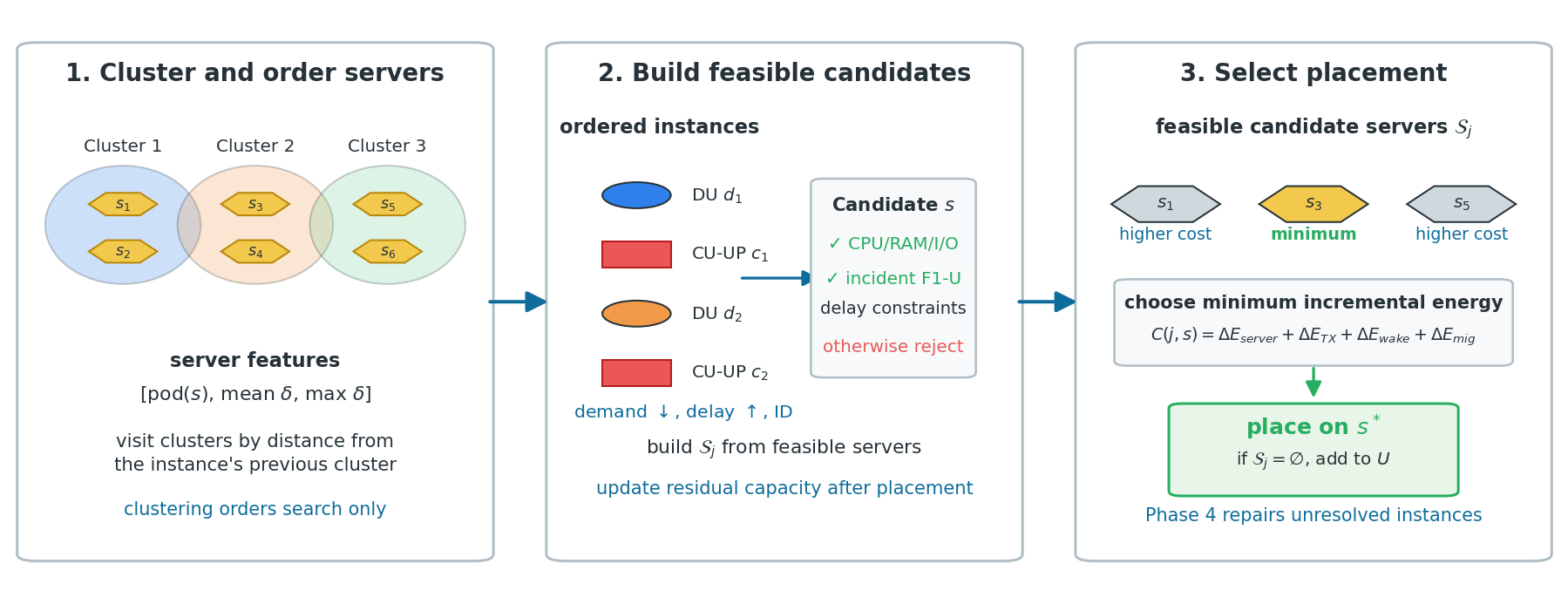}
  \caption{Illustration of the Phase-2 delay-aware physical mapping procedure. Server clustering determines the search order only. For each DU or CU-UP instance, servers satisfying the residual-resource and currently determined F1-U-delay constraints form the candidate set; the minimum incremental-energy candidate is selected, or the instance is added to the unresolved set for Phase-4 repair.}
  \label{fig:map_du_cu_to_servers}
\end{figure*}

Phase 2 transforms the logical assignment $(q,z)$ into a physical placement.
To guide this process, server clustering determines the search order but does not determine placement feasibility.
For each instance, a candidate placement is accepted only if it satisfies the residual {\CPU}, {\RAM}, and {\IO} capacities and the static one-way delay constraints.
Among the feasible candidates, the algorithm evaluates the same kWh components as Eq.~\eqref{eq:total_objective}, including the fixed wake-up charge for activating an inactive server and the precomputed migration cost when an instance leaves its previous server.
If no feasible candidate exists, the instance is added to $\mathcal{U}$ and deferred to the Phase 4 repair procedure.
After the initial mapping, Phase 3 attempts to consolidate the placement.
Because fixed idle-power savings are realized only after the final instance leaves a source server, the algorithm evaluates complete server-evacuation plans rather than isolated moves.
A plan is accepted only after its wake-up and migration costs have been included in the total-energy change.


Phase 3 evaluates candidate moves sequentially using the same total-energy objective as the MILP.
The working placement is initialized from the placement obtained in the previous control interval.
An accepted move updates only the selected instance and is carried forward to subsequent evaluations, whereas a rejected move leaves the working placement unchanged.
Migration energy is represented by the precomputed kWh coefficient in Eq.~\eqref{eq:migration_coefficient}; no dimensionless weighting factor is used.

\begin{algorithm}[h]
  \small
  \caption{Phase 3: MigrateSequential (Common-Unit Validation)}
  \label{alg:migrate_energy_aware}

  \begin{algorithmic}[1]
    \Statex \textbf{Input:} Candidate placement $x^*$, previous placement $\PreviousPlacement$, assignment $q$, pairing $z$,
    \Statex \hspace*{1.cm} migration coefficients $\MigrationCoeff$, wake costs $E^{\mathrm{wake}}$, $\epsilon$, $u_{\mathrm{cons}}$
    \Statex \textbf{Output:} Final placement $x$, migration set $m$
    \Statex
    \State $x\gets\PreviousPlacement$ and $m\gets\emptyset$
    \State Build changed-instance list $\mathcal{M}\gets\{j:x_j^*\neq \PreviousPlacement_j\}$
    \State Order $\mathcal{M}$ by increasing estimated $\Delta E(j,x_j^*)$, then instance ID
    \For{each $j\in\mathcal{M}$}
    \State Construct $x'$ by moving only $j$ from its current server in $x$ to its target in $x^*$
    \State Recompute affected server, transmission, wake-up, and migration energy terms
    \State $\Delta E\gets E^{\mathrm{total}}(x',q,z)-E^{\mathrm{total}}(x,q,z)$
    \If{$x'$ satisfies assignment, capacity, and delay constraints \textbf{and} $\Delta E\le-\epsilon$}
    \State $x\gets x'$ and $m\gets m\cup\{j\}$
    \EndIf
    \EndFor
    \For{each active server $\ServerIdx$ with $\Utilization_{\ServerIdx}\le u_{\mathrm{cons}}$}
    \State Build a deterministic feasible evacuation plan $x'$ for all instances on $\ServerIdx$
    \If{the plan exists and $E^{\mathrm{total}}(x',q,z)\le E^{\mathrm{total}}(x,q,z)-\epsilon$}
    \State Accept the complete plan and update $m$
    \EndIf
    \EndFor
    \Statex \textbf{Return} $(x, m)$
  \end{algorithmic}
\end{algorithm}
%


Finally, in Phase 4, Algorithm~\ref{alg:repair_feasibility} attempts to repair residual feasibility violations after migration and computes the violation statistics reported in the experimental evaluation.

\begin{algorithm}[h]
  \small
  \caption{Phase 4: RepairFeasibility (Post-Processing Constraint Repair)}
  \label{alg:repair_feasibility}

  \begin{algorithmic}[1]
    \Statex \textbf{Input:} Placement $x$, assignment $q$, pairings $z$, migrations $m$, unassigned set $\mathcal{U}$, budget $I_{\mathrm{rep}}$
    \Statex \textbf{Output:} Repaired $(x,q,z,m)$ and violation report $\nu$
    \Statex
    \State Compute $\Phi(x,q,z)=(|\mathcal{U}|,v_{\mathrm{assign}},v_{\mathrm{cap}},v_{\mathrm{delay}},v_{\mathrm{conn}})$
    \State Store pre-repair statistics $(r_{\mathrm{pre}}, \bar{v}_{\mathrm{pre}}, v^{\max}_{\mathrm{pre}})$
    \For{$t = 1$ to $I_{\mathrm{rep}}$}
    \If{$\Phi(x,q,z)=\mathbf{0}$}
    \State \textbf{break}
    \EndIf
    \State Build deterministic candidates: place one unassigned instance, reassign one slice-flow CU-UP, move one instance, or roll back one migration
    \State Keep only moves that strictly decrease $\Phi$ in lexicographic order
    \If{no improving move exists}
    \State \textbf{break}
    \EndIf
    \State Apply the improving move with minimum energy increase
    \State Recompute $\Phi(x,q,z)$
    \EndFor
    \State Store post-repair statistics $(r_{\mathrm{post}}, \bar{v}_{\mathrm{post}}, v^{\max}_{\mathrm{post}})$ and repaired-instance count $n_{\mathrm{rep}}$
    \State $\nu \gets (r_{\mathrm{pre}}, r_{\mathrm{post}}, \bar{v}_{\mathrm{post}}, v^{\max}_{\mathrm{post}}, n_{\mathrm{rep}})$
    \Statex \textbf{Return} $(x,q,z,m,\nu)$
  \end{algorithmic}
\end{algorithm}

\subsection{Feasibility Repair: Termination and Limitations}

The state space defined by $(x,q,z,m)$ is finite. At each repair step, Phase 4 accepts only a move that strictly decreases the violation vector $\Phi$ in lexicographic order; therefore, it cannot cycle among previously visited violation states. Moreover, the search is limited to at most $I_{\mathrm{rep}}$ iterations. Thus, the repair procedure always terminates.
However, termination does not guarantee feasibility. Phase 4 is a bounded greedy local search and may stop when no candidate move reduces $\Phi$, even though a different sequence of moves could reach a feasible MILP solution. We therefore claim neither global optimality nor completeness with respect to the considered neighborhood. If $\Phi\neq\mathbf{0}$ at termination, the interval is classified as infeasible by the heuristic and excluded from feasible-energy comparisons.
Accordingly, for each time interval, we report the violation rate before and after repair, the mean and maximum post-repair violation magnitudes, the number of repaired instances, and the final feasibility status.

\subsection{Complexity Analysis}

Let $N_{\mathrm{sf}}=\sum_{\FlowIdx}|\SliceSet_{\FlowIdx}|$, $J=|\DUSet|+|\CUSet|$, and let $M_{\mathrm{rep}}$ and $C_{\mathrm{check}}$ denote the number of repair candidates and the cost of one full feasibility check. With fixed clustering restarts and iteration limits, a conservative bound is
\begin{equation*}
  \begin{aligned}
    \mathcal{O}\!\big(&R_k I_k N_{\mathrm{sf}} k_{\mathrm{pair}}
      +|\CUSet|^3+I_k|\ServerSet||\mathrm{pods}|\\
      &+(J+N_{\mathrm{sf}}|\CUSet|)|\ServerSet|^2
    +I_{\mathrm{rep}}M_{\mathrm{rep}}C_{\mathrm{check}}\big).
  \end{aligned}
\end{equation*}

Each term corresponds to one phase of our heuristic algorithm:

\begin{itemize}
  \item \textbf{Phase 1:} $R_k$ restarts of at most $I_k$ $k$-means iterations cost $O(R_k I_k N_{\mathrm{sf}}k_{\mathrm{pair}})$ for a fixed feature dimension. Hungarian matching contributes $O(|\CUSet|^3)$ and is retained explicitly in the bound.

  \item \textbf{Phase 2:} Server clustering costs $O(I_k|\ServerSet||\mathrm{pods}|)$. Ordered placement and CU-UP-indexed delay checks contribute at most $O((J+N_{\mathrm{sf}}|\CUSet|)|\ServerSet|^2)$.

  \item \textbf{Phase 3:} At most $J$ changed instances and at most $|\ServerSet|$ evacuation plans are evaluated. Incremental caches reduce practical cost, while the displayed bound conservatively permits full affected-route checks.

  \item \textbf{Phase 4:} The explicit term $I_{\mathrm{rep}}M_{\mathrm{rep}}C_{\mathrm{check}}$ avoids assuming that candidate generation or feasibility checking is negligible.
\end{itemize}

Regarding space complexity, the algorithm’s memory footprint scales as follows:
\begin{equation}
  O(N_{\mathrm{sf}}|\CUSet|+|\ServerSet|^2+J|\ServerSet|+|\DUSet||\CUSet|) \nonumber
\end{equation}

Memory is dominated by the assignment matrix $q$, the DU-to-CU-UP pairing matrix $z$, the inter-server delay matrix, placement/residual-capacity arrays, and the bounded Phase-4 candidate list. This asymptotic analysis establishes polynomial growth for fixed $R_k$, $I_k$, and $I_{\mathrm{rep}}$, but does not by itself quantify the runtime advantage over the MILP.


\section{Experiments}

We use a 24-hour traffic trace that is time varying across hours but deterministic for each hourly optimization.
At the beginning of hour $t$, each method receives the exact preprocessed demand for hour $t$ and the placement inherited from hour $t-1$; demand is held fixed during that hourly solve, and the demand values for hours $t+1,t+2,\ldots$ are unavailable.
Thus, \emph{time varying} refers to the observed hour-to-hour change in traffic, whereas \emph{deterministic} means that no demand uncertainty, probability distribution, or stochastic scenario is included in an individual optimization run.


\subsection{5G traffic and edge cloud setup}
\label{sec:traffic_setup}

\begin{figure}[pos=htbp]
  \centering
  \includegraphics[width=1\linewidth]{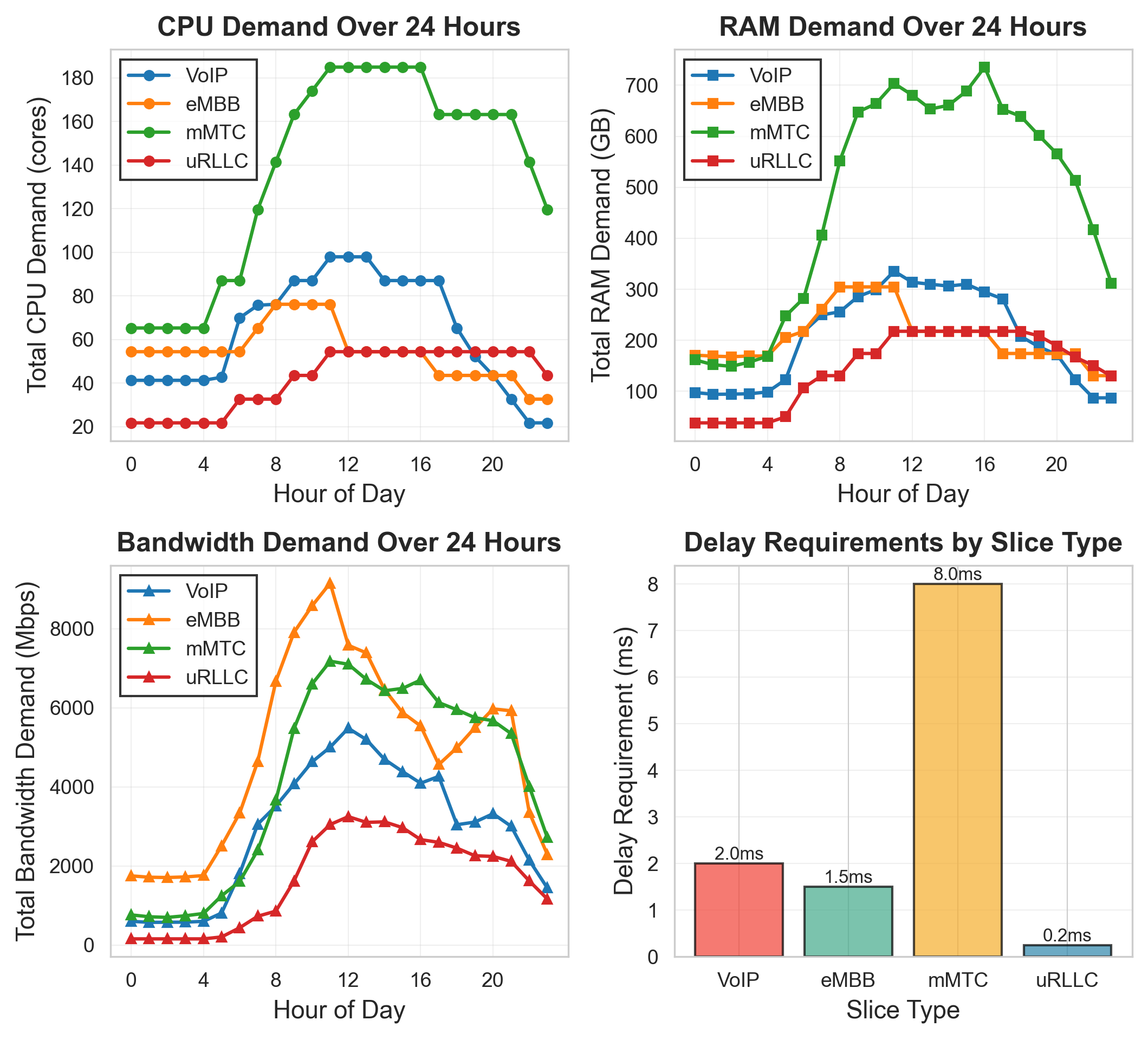}
  \caption{Twenty-four-hour traffic demand by slice type in one edge cloud.}
  \label{fig:traffic_demands_by_slice_type}
\end{figure}

To evaluate the MILP and heuristic under realistic operating conditions, we generate hourly demand profiles from Montreal traffic traces following~\cite{delgado2022network} and simulate the edge-cloud environment using CloudSim~\cite{calheiros2011cloudsim}.
As shown in Fig.~\ref{fig:traffic_demands_by_slice_type}, the dataset captures hourly variations across representative 5G slice types: eMBB, uRLLC, mMTC, and VoIP, which is treated as a uRLLC-type service with minor flexibility depending on network conditions.
These profiles drive the hourly placement and migration decisions.

The slice traffic properties and baseline resource profiles are adapted from the 5G network-simulation setup in our previous work~\cite{tran2025proactive}, with the scope restricted to the DU-to-CU-UP segment considered here.
Table~\ref{tab:slice_characteristics} summarizes these characteristics and the experimental F1-U delay budgets.
In this table, one \emph{demand unit} denotes one active slice-flow demand item of the indicated slice generated by preprocessing for one control interval; it is a workload-accounting item, not one UE, packet, Mbps, or CNF instance.
Each demand unit contributes the listed baseline reservation to the resource vector: vCPU, RAM in gigabytes (GB), and server-NIC I/O rate in Mbps.
For multiple demand units, these resource vectors are summed, after which the external scaling stage maps the aggregate demand to vertically sized or horizontally replicated DU/CU-UP CNF instances.
This request-vector abstraction is consistent with resource-aware NFV studies that represent time-varying VNF and service-chain requirements using CPU, memory, and network-bandwidth dimensions~\cite{tang2019dynamic}.
In our study, the network-bandwidth dimension is labelled I/O because it is enforced against the server NIC capacity; the qualitative ``Bandwidth Demand'' column instead describes the relative traffic intensity of each slice.
\begin{table*}[pos=ht]
  \centering
  \caption{Network slice characteristics, per-demand-unit resource profiles, and experimental F1-U delay budgets}
  \label{tab:slice_characteristics}
  \scriptsize
  \begin{tabular}{l l l l l l l}
    \toprule
    \textbf{Slice} & \textbf{Traffic Pattern} & \makecell{\textbf{F1-U delay}\\\textbf{budget}} & \textbf{Bandwidth Demand} & \makecell{\textbf{CPU/demand}\\\textbf{unit}} & \makecell{\textbf{RAM/demand}\\\textbf{unit}} & \makecell{\textbf{I/O/demand}\\\textbf{unit}} \\
    \midrule
    eMBB  & High with pronounced daily peaks & 1.5 ms & Very high  & 0.5 vCPU & 1.0 GB & 12 Mbps \\
    uRLLC & Bursty and delay-critical      & 0.5 ms & High       & 0.8 vCPU & 1.2 GB & 5 Mbps  \\
    mMTC  & Many small packets             & 8 ms & Very low  & 0.2 vCPU & 0.3 GB & 1.5 Mbps \\
    VoIP  & Diurnal, small fluctuations    & 2 ms   & Low--medium & 0.4 vCPU & 0.6 GB & 2 Mbps \\
    \bottomrule
  \end{tabular}
\end{table*}
The budgets in Table~\ref{tab:slice_characteristics} apply only to the modelled one-way DU-to-CU-UP/F1-U path and are not standardized end-to-end service requirements.
They are conservative internal modelling choices whose scale and ordering are motivated by peer-reviewed RAN-placement and slicing studies reporting sub-millisecond uRLLC limits, millisecond-scale functional-split and VoIP bounds, and looser eMBB and mMTC tolerances~\cite{zorello2022baseband,klinkowski2020flowallocation,tseliou2019netslic}.
In the trace, eMBB has high bandwidth demand and moderate compute requirements, uRLLC has the strictest delay budget and higher per-demand-unit CPU/RAM demand, mMTC has low per-demand-unit resource demand, and VoIP has modest resource requirements and limited fluctuations.

To generate the optimization inputs under these hour-to-hour traffic variations, an external preprocessing step first applies vertical scaling to each instance based on its current load.
Notably, a single DU can serve multiple network-slice traffic types simultaneously.
Each instance is scaled vertically up to a predefined threshold (e.g., 30\% of a server’s total capacity). Once this limit is reached, horizontal scaling is triggered.
When traffic decreases, the scaling direction is reversed accordingly.
Thus, the system prioritizes horizontal scale-in to reduce the number of active instances, followed by vertical scale-down to further consolidate resource usage.
As a result, at each hour, preprocessing produces sets of DUs and CU-UPs that correspond to the traffic demand, along with the traffic-flow mapping to the associated DUs.
The number of active containers followed the traffic demand curve, peaking at 50 instances at hour 11.
These sets are then used as fixed inputs to our MILP model and heuristic; the scaling procedure itself is not optimized by either method.

%
\begin{table}[pos=ht]
  \centering
  \caption{Homogeneous edge-cloud server configuration}
  \begin{tabular}{ll}
    \toprule
    \textbf{Parameter}            & \textbf{Value}      \\
    \midrule
    Number of Pods                & 3                   \\
    Servers per Pod               & 4                   \\
    Total Servers                 & 12                  \\
    \midrule
    CPU Capacity (per server)     & 64 vCPU           \\
    RAM Capacity (per server)     & 256 GB            \\
    Bandwidth Capacity (per server) & 25 Gbps  \\
    Storage Capacity (per server) & 1000 GB           \\
    \bottomrule
  \end{tabular}
  \label{tab:fat_tree_edge_cloud}
\end{table}

\begin{figure}
  \centering
  \includegraphics[width=0.8\linewidth]{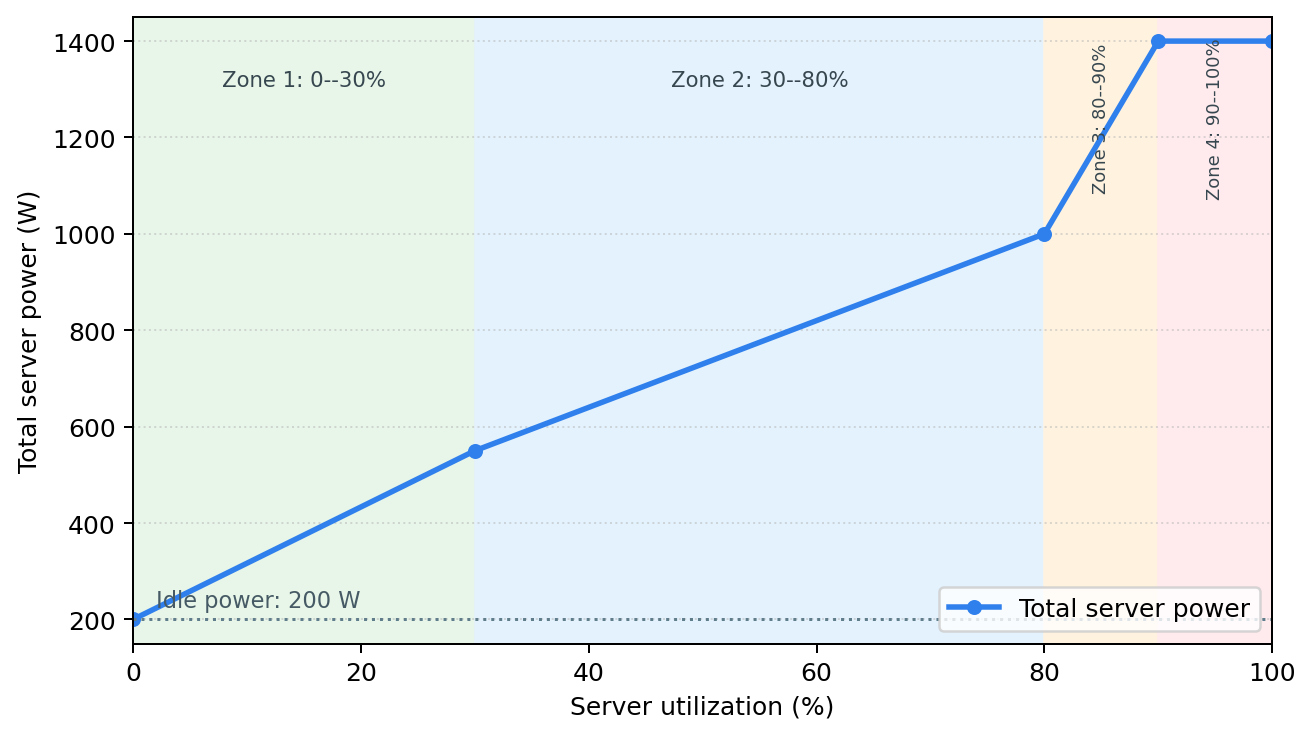}
  \caption{Four-zone total server-power curve used to derive load-dependent incremental power.}
  \label{fig:server_power_zone}
\end{figure}

We simulate a three-pod edge-cloud topology using 12 homogeneous logical servers (see Table~\ref{tab:fat_tree_edge_cloud}); the server-to-server delay matrix is generated from the switch paths described in Table~\ref{tab:delay_fat_tree}.
The power curve is calibrated to one Dell PowerEdge XR8720t-class server profile rather than mixing cloud-instance and physical-server power models. Figure~\ref{fig:server_power_zone} shows total server power; the MILP subtracts the 200~W idle value and models only the remaining load-dependent incremental power in $\PiwisePower_{\ServerIdx}$.
The fitted total-power breakpoints are $(0,200)$, $(0.30,550)$, $(0.80,1000)$, $(0.90,1400)$, and $(1.00,1400)$, where utilization is a fraction and power is in watts. Subtracting 200~W gives the incremental coefficients in Table~\ref{tab:incremental_power_coefficients}. RAM and I/O affect the utilization variable through $w_r$.
Since the nominal experiments impose $\MaxUtil=0.70$, only the portions of Zones~1 and~2 up to that cap are reachable; Zones~3 and~4 are retained for utilization-cap sensitivity experiments.

\begin{table}[pos=ht]
  \centering
  \caption{Load-dependent incremental-power coefficients}
  \label{tab:incremental_power_coefficients}
  \scriptsize
  \begin{tabular}{cccc}
    \toprule
    Zone & Utilization interval & $a_i$ (W) & $b_i$ (W) \\
    \midrule
    1 & $[0,0.30]$ & $1166.7$ & $0$ \\
    2 & $(0.30,0.80]$ & $900$ & $80$ \\
    3 & $(0.80,0.90]$ & $4000$ & $-2400$ \\
    4 & $(0.90,1.00]$ & $0$ & $1200$ \\
    \bottomrule
  \end{tabular}
\end{table}
For delay modelling, the one-way intra-edge-cloud network delay is obtained from the placement-dependent components in Table~\ref{tab:delay_fat_tree}. The constrained DU-to-CU-UP/F1-U delay is the sum of this network delay and the fixed F1-U processing component reported in the same table.

\begin{table}[pos=ht]
  \centering
  \caption{Modelled one-way DU-to-CU-UP/F1-U delay components}
  \scriptsize
  \begin{tabularx}{\columnwidth}{>{\raggedright\arraybackslash}X>{\raggedright\arraybackslash}X}
    \toprule
    \textbf{Parameter}     & \textbf{Value / Formula}           \\
    \midrule
    Same-server network delay & 0.05 ms \\
    Base link delay        & 0.05 ms (per one-way link)                 \\
    ToR switch processing         & 0.1 ms                            \\
    Aggregation switch processing         & 0.15 ms                           \\
    Fat-tree core-switch processing        & 0.2 ms                            \\
    Fixed F1 processing          & 0.1 ms (CU side) + 0.1 ms (DU side) = 0.2 ms \\

    \bottomrule
  \end{tabularx}
  \label{tab:delay_fat_tree}
\end{table}

\subsection{Experimental parameter settings}

To ensure reproducibility, we fix the MILP and heuristic settings shown in Table~\ref{tab:exp_parameter_settings} across all experiments unless stated otherwise.
The proposed MILP is implemented using Gurobi Optimizer, with the relative MIP-gap target set to $10^{-3}$.
Most instances reached the solver's optimal status; for the remaining instances, the reported MILP value is the best feasible solution obtained within the time limit.
The heuristic uses fixed clustering restarts, iteration limits, and pseudo-random seed.
The same parameter set is used for both Single-CU and Multi-CU scenarios, except for the scenario-specific value of $\MaxCUPerDU$.

\begin{table}[pos=t]
  \centering
  \caption{Fixed experimental inputs and hyperparameters for MILP and heuristic runs}
  \label{tab:exp_parameter_settings}
  \scriptsize
  \begin{tabularx}{\columnwidth}{>{\raggedright\arraybackslash}m{0.36\columnwidth}X}
    \toprule
    \textbf{Parameter} & \textbf{Setting used in experiments} \\
    \midrule
    $\TrafficFlowData$ & Montreal hourly traffic profiles generated following~\cite{delgado2022network} \\
    $\ControlInterval$ & $1$ hour \\
    $P_{\ServerIdx}^{\idle},P_{\ServerIdx}^{\mathrm{inc,max}}$ & $200$~W and $1200$~W; coefficients in Table~\ref{tab:incremental_power_coefficients} \\
    $E_{\ServerIdx}^{\mathrm{wake}}$ & Fixed $0.01$~kWh per off-to-on transition (scenario assumption; sensitivity required) \\
    $B_j^{\mathrm{mig}}$ & $1.2$ times the allocated RAM footprint of instance $j$ \\
    $\beta_{\ServerIdx\ServerIdx'}^{\mathrm{mig}}$ & Minimum of $10$~Gbps and residual migration-path bandwidth \\
    $P_{\ServerIdx}^{\mathrm{src}},P_{\ServerIdx'}^{\mathrm{dst}}$ & $0.05P^{\mathrm{inc,max}}$ at each endpoint \\
    $\epsilon_{\ServerIdx\ServerIdx'}^{\mathrm{net}}$ & Path-specific CloudSim transmission-energy coefficient (kWh/GB) \\
    MILP implementation (Gurobi) & Relative MIP gap $10^{-3}$ \\
    $\MaxCUPerDU$ (Single-CU: one CU-UP) & $1$ \\
    $\MaxCUPerDU$ (Multi-CU: CU-UP bound) & $3$ \\
    $\eta_{\CU/\DU}$ & $2$ \\
    $R_k,I_k,\xi$ & $10$ restarts, $100$ iterations, seed $42$ \\
    $I_{\mathrm{rep}}$ & $15$ \\
    $\epsilon$ & $10^{-4}$~kWh \\
    $u_{\mathrm{cons}}$ & $0.30$; consolidation candidates satisfy $\Utilization_{\ServerIdx} \leq u_{\mathrm{cons}}$ \\
    \bottomrule
  \end{tabularx}
\end{table}


\subsection{Numerical results}

We compare the Single-CU and Multi-CU configurations, which permit one or multiple CU-UP targets per DU, respectively.
To isolate the value of jointly optimizing placement and migration, we compare the proposed formulation with a migration-unaware per-interval placement baseline.
At each hour, this baseline optimizes the current placement without terms for migration energy or placement continuity.
It does not force instances to move; an instance may remain on its previous server, and only placement changes selected by the optimizer are subsequently counted as induced migrations.
For brevity, the figures label this baseline \textit{Placement-Only}.
We also compare the $k$-means-based heuristic with the MILP and common placement baselines.

\subsubsection{MILP performance: Single-CU and Multi-CU scenarios}

Migration energy is computed from Eqs.~\eqref{eq:migration_duration}--\eqref{eq:migration_coefficient}, and wake-up energy is charged through Eqs.~\eqref{eq:wake_lb}--\eqref{eq:wake_ub}.
Consequently, all objective components are expressed in kWh.

\begin{figure*}[pos=ht!p]
  \centering
  \begin{subfigure}[b]{0.32\linewidth}
    \centering
    \includegraphics[width=\linewidth]{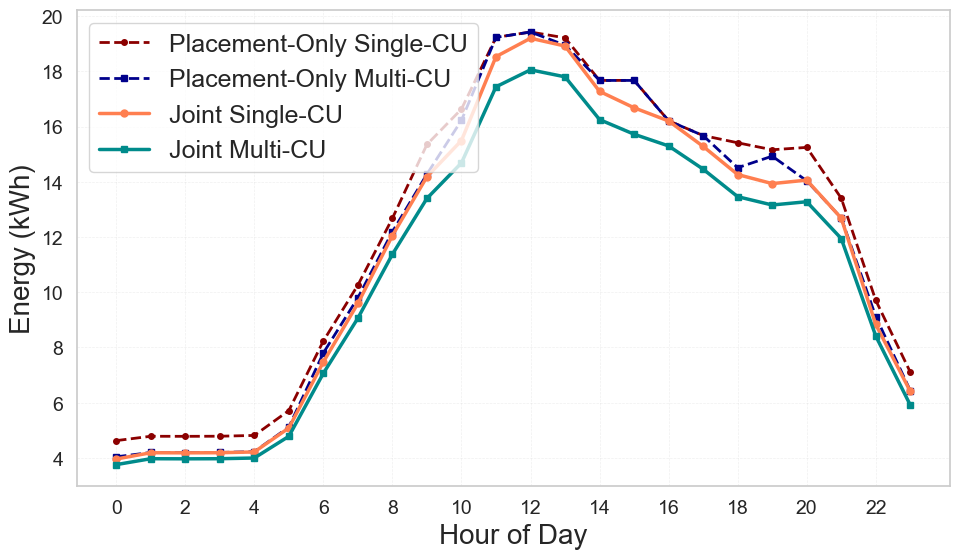}
    \caption{Hourly total energy.}
    \label{fig:ilp_total_energy}
  \end{subfigure}
  \hfill 
  \begin{subfigure}[b]{0.32\linewidth}
    \centering
    \includegraphics[width=\linewidth]{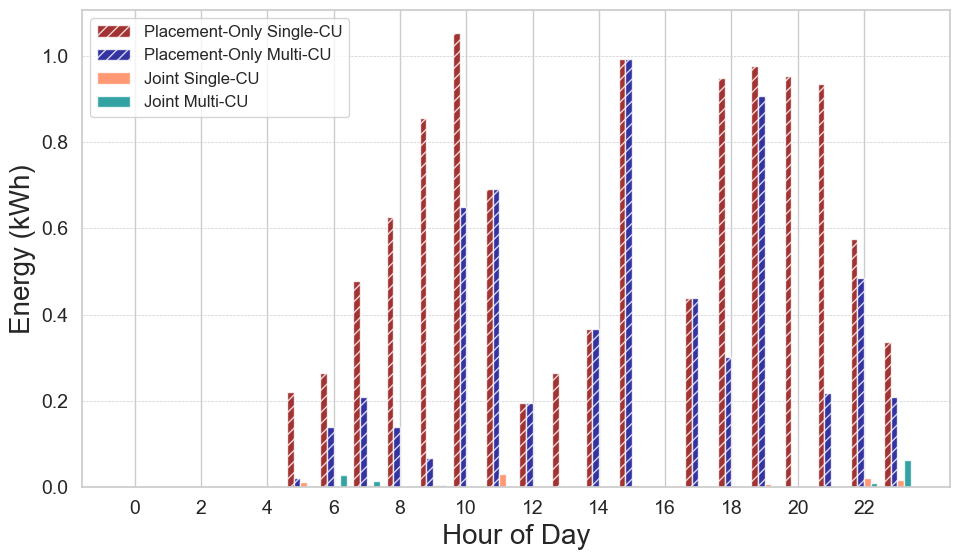}
    \caption{Hourly migration energy.}
    \label{fig:ilp_migration_energy}
  \end{subfigure}
  \hfill 
  \begin{subfigure}[b]{0.32\linewidth}
    \centering
    \includegraphics[width=\linewidth]{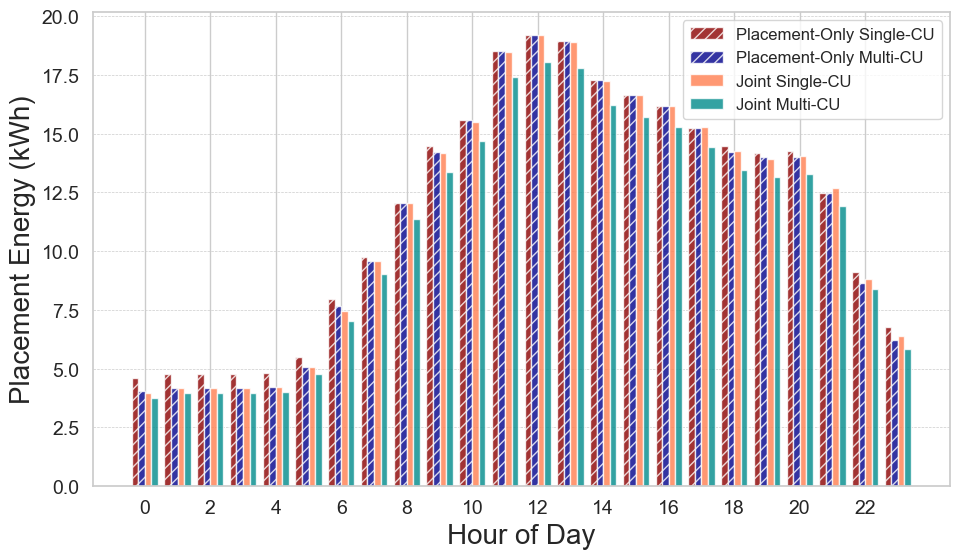}
    \caption{Hourly placement energy.}
    \label{fig:ilp_placement_energy}
  \end{subfigure}

  \caption{Twenty-four-hour MILP energy breakdown for migration-unaware per-interval placement and joint placement and migration. The legend abbreviates the migration-unaware baseline as \textit{Placement-Only}; this baseline permits unchanged placements but does not include migration energy or placement continuity in its objective.}
  \label{fig:ilp_energy_breakdown}
\end{figure*}

\begin{figure}[pos=ht]
  \centering
  \includegraphics[width=0.9\linewidth]{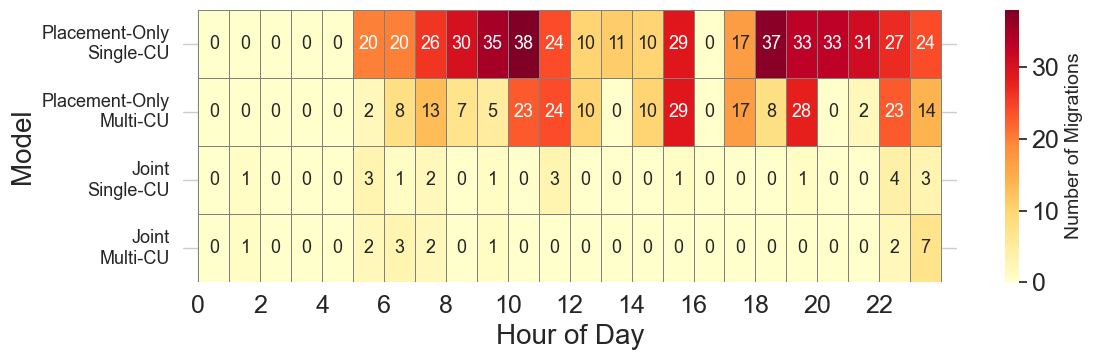}
  \caption{Hourly number of migrations.}
  \label{fig:ilp_num_migration}
\end{figure}

Fig.~\ref{fig:ilp_total_energy} shows that joint placement-and-migration optimization consumes less total energy than the migration-unaware baseline in this trace.
Over the 24-hour period, joint Multi-CU consumes 5.7\% less energy than joint Single-CU and also remains below both migration-unaware configurations.
During low-traffic hours 0--4, the average hourly energy difference between the two joint configurations is only 0.99~kWh (4.39\%), with Multi-CU consuming less energy.
During peak traffic, this difference increases to 2.11~kWh (8.3\%), indicating that slice-specific CU-UP assignment is most beneficial when capacity is constrained.
By contrast, the energy difference between the migration-unaware Single-CU and Multi-CU configurations narrows during peak traffic because high utilization limits placement flexibility.
As shown in Fig.~\ref{fig:ilp_num_migration}, joint Single-CU triggers substantially more migrations than joint Multi-CU, particularly during rapid load fluctuations.
The Single-CU constraint provides fewer feasible placement choices, causing the optimizer to rely more heavily on migration to satisfy instantaneous load and delay requirements.
The larger Multi-CU decision space reduces both the number of migrations and the corresponding migration energy, as shown in Fig.~\ref{fig:ilp_migration_energy}.
The migration-unaware baseline also produces more induced migrations because it selects target servers without minimizing migration energy or rewarding placement continuity.
In contrast, joint optimization discourages a relocation when its energy cost exceeds its placement benefit.

\begin{figure}[pos=ht]
  \centering
  \includegraphics[width=1\linewidth]{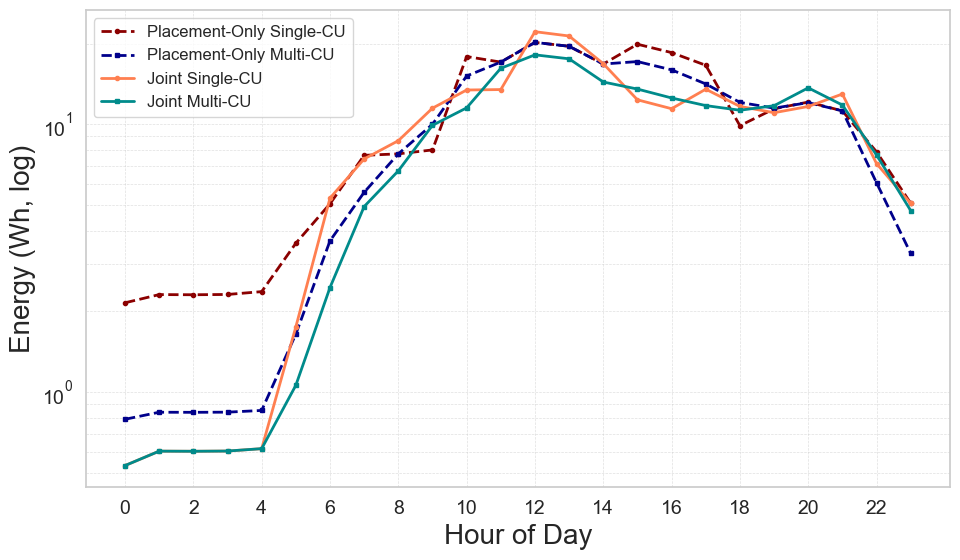}
  \caption{Transmission energy for the four MILP configurations (logarithmic scale).}
  \label{fig:ilp_transmission_energy}
\end{figure}

As reported in Fig.~\ref{fig:ilp_transmission_energy}, the migration-unaware Multi-CU configuration consumes less transmission energy during low-traffic periods, whereas joint Multi-CU consumes slightly more during some peak hours.
This increase reflects a trade-off between longer internal paths and lower server energy and is selected only when it reduces total energy.
For the two joint configurations, transmission energy is nearly identical during hours 0--4, and the average relative difference over the remaining hours is 3\%.
Moreover, Fig.~\ref{fig:ilp_placement_energy} shows that Multi-CU consistently achieves lower placement energy because slice-aware mappings provide greater flexibility in distributing slice-flow groups across DU/CU-UP instances and servers.
At the highest traffic levels in the migration-unaware setting, the energy difference becomes negligible because resource saturation causes both configurations to converge toward similar feasible placements.
Over 24 hours, placement dominates total energy, followed by migration, while transmission contributes only 0.07\%.

\begin{figure}[pos=ht]
  \centering
  \includegraphics[width=1\linewidth]{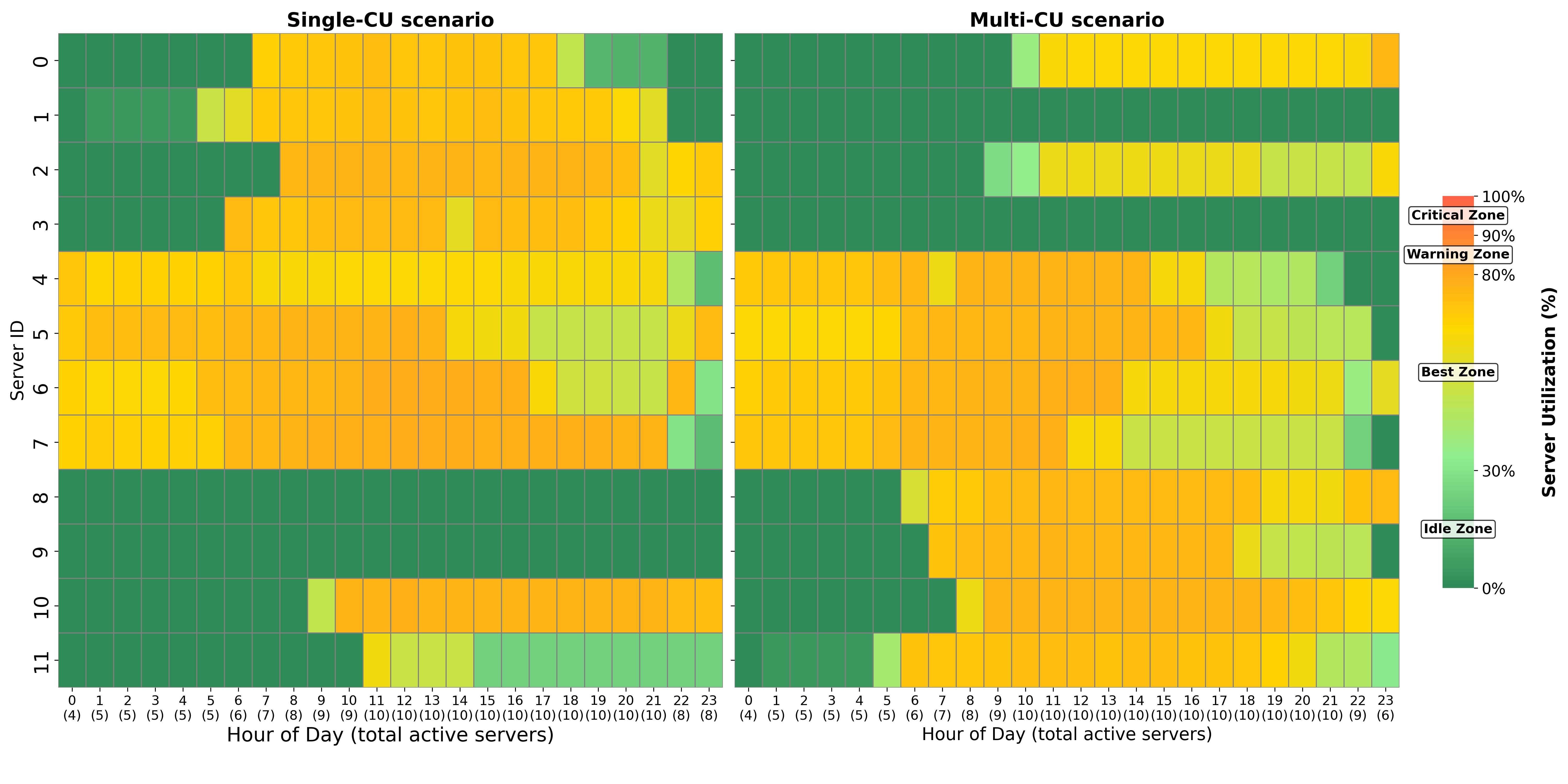}
  \caption{Server load distribution across servers (joint placement and migration approach).}
  \label{fig:ilp_server_load}
\end{figure}

Fig.~\ref{fig:ilp_server_load} further shows that both joint configurations activate the same number of servers during low-traffic hours 0--7.
During peak hours 8--16, Multi-CU distributes the load more evenly and causes fewer servers to reach the warning-zone utilization threshold.
During hours 17--23, it occasionally activates an additional server, trading higher fixed server energy for lower load-dependent power on heavily loaded servers.
By contrast, Single-CU pushes several servers into the warning zone, for example at hour 21, resulting in higher energy consumption.


\subsubsection{Heuristic performance}


\begin{figure*}[pos=htp]
  \centering
  \begin{subfigure}[b]{0.32\linewidth}
    \centering
    \includegraphics[width=\linewidth]{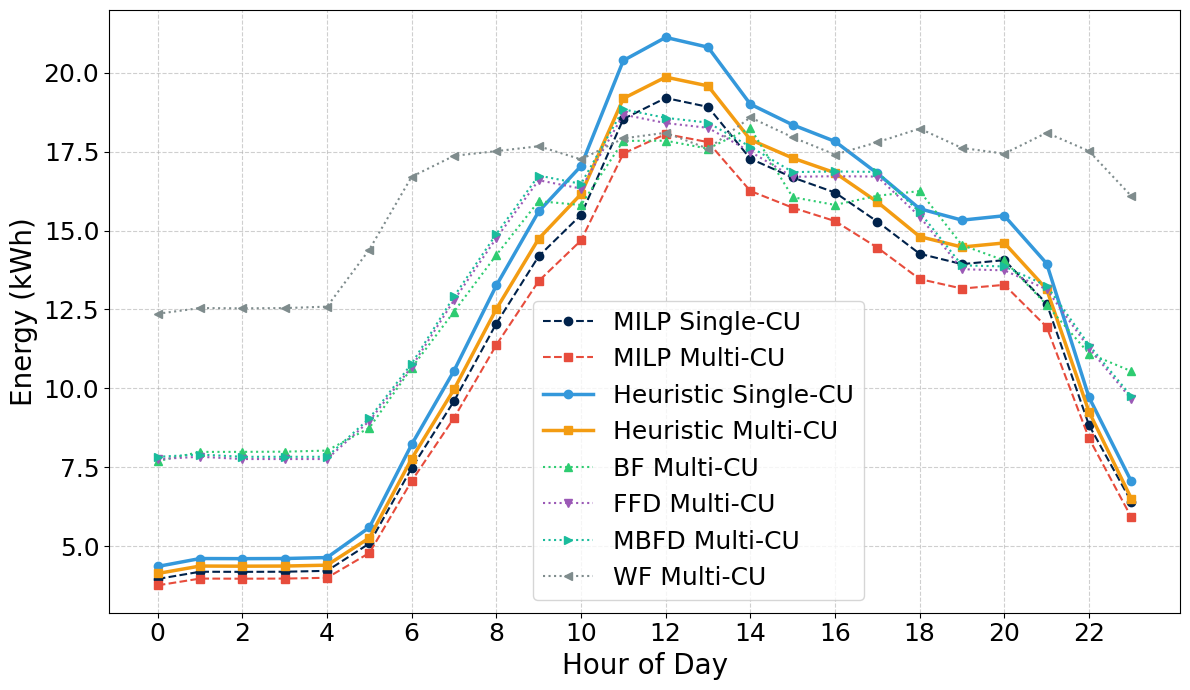}
    \caption{Hourly total energy.}
    \label{fig:hourly_energy_ilp_vs_heuristic}
  \end{subfigure}
  \hfill
  \begin{subfigure}[b]{0.32\linewidth}
    \centering
    \includegraphics[width=\linewidth]{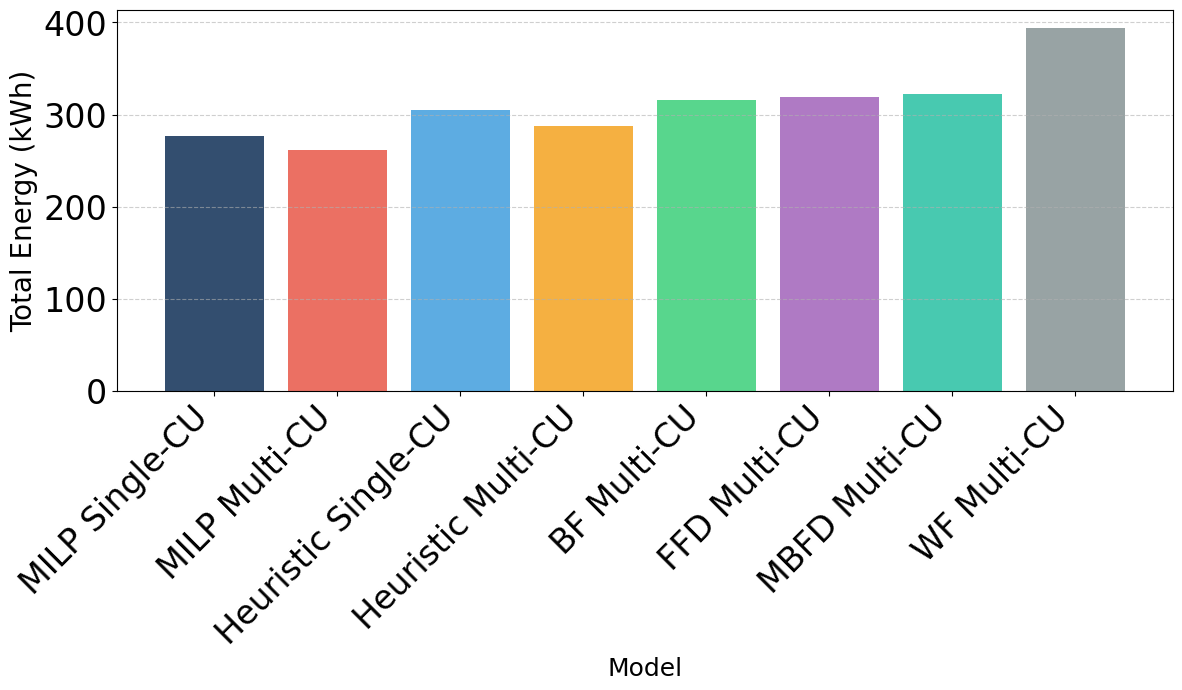}
    \caption{Twenty-four-hour total energy.}
    \label{fig:heuristic_total_energy_day}
  \end{subfigure}
  \hfill
  \begin{subfigure}[b]{0.32\linewidth}
    \centering
    \includegraphics[width=\linewidth]{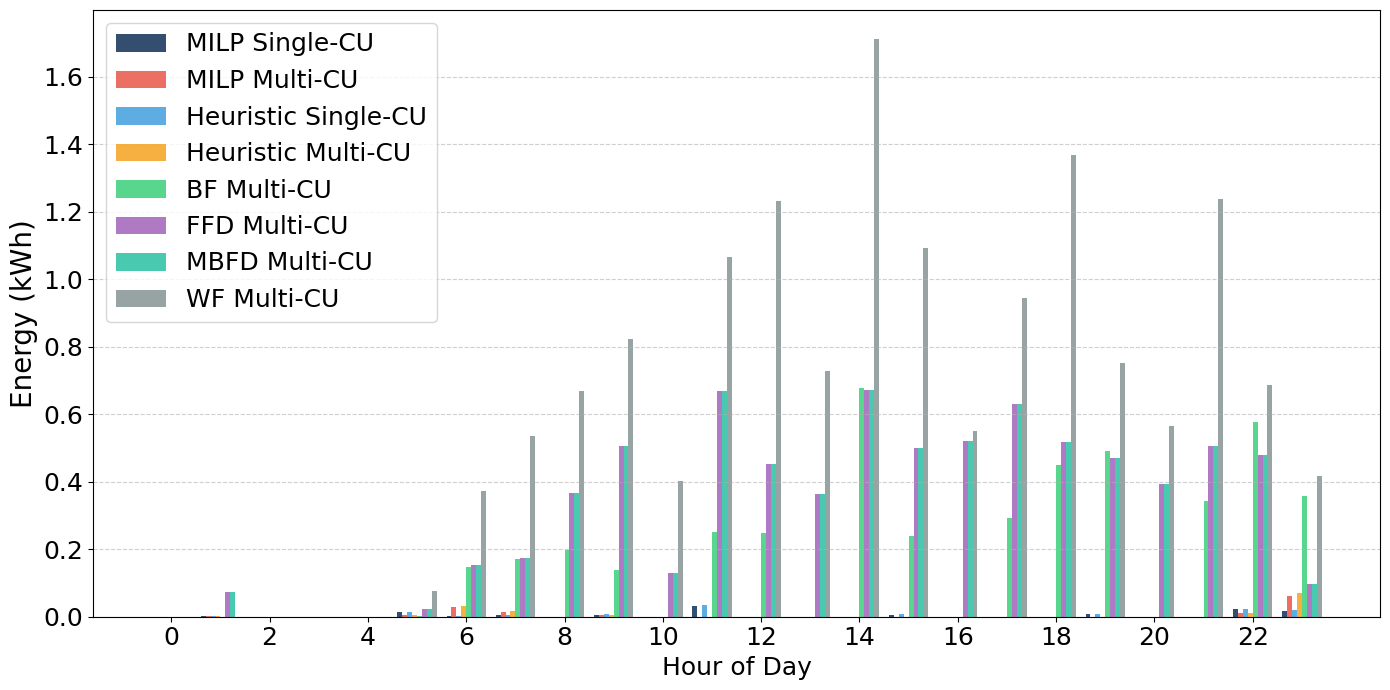}
    \caption{Hourly migration energy.}
    \label{fig:heuristic_migration_energy}
  \end{subfigure}
  \caption{Energy comparison over the 24-hour workload for the MILP, proposed heuristic, and cloud-placement baselines: (a) hourly total energy, (b) 24-hour total energy, and (c) hourly migration energy.}
  \label{fig:heuristic_energy_comparison}
\end{figure*}

We evaluate the proposed $k$-means-based heuristic (H\_EJPM) on the same 24-hour O-RAN workload against two reference groups.
First, we compare H\_EJPM with the proposed joint placement-and-migration MILP under both the Single-CU and Multi-CU association scenarios.
Second, under the more energy-efficient Multi-CU-UP scenario, we compare H\_EJPM with common cloud-placement baselines: Best Fit (BF)~\cite{shirvastava2017best}, First Fit Decreasing (FFD), Modified Best Fit Decreasing (MBFD)~\cite{moges2019energy}, and Worst Fit (WF)~\cite{dhahbi2021load}.
The selected baselines represent distinct placement behaviours: BF and FFD emphasize compact packing and server consolidation, MBFD applies an energy-aware decreasing-order policy, and WF spreads the workload across available servers.
Together, these comparisons assess how closely H\_EJPM approaches the MILP's energy performance and whether it improves on standard placement rules.

Fig.~\ref{fig:heuristic_energy_comparison} summarizes the hourly total energy, 24-hour total energy, and hourly migration energy for all evaluated methods.
In the Multi-CU scenario, the proposed heuristic consumes 286.4~kWh, compared with 261.2~kWh for the MILP; in the Single-CU scenario, the corresponding values are 304.6 and 276.9~kWh, respectively.
Thus, relative to the MILP, the heuristic has energy gaps of approximately 9.7\% in the Multi-CU scenario and 10.0\% in the Single-CU scenario.
These gaps are attributable almost entirely to placement energy; migration and transmission energy are negligible for both methods.
The difference primarily reflects weaker server consolidation, as the heuristic activates an average of 9 servers compared with 8 for the MILP, rather than excessive remapping.

Importantly, the heuristic preserves the relative benefit of Multi-CU operation.
Moving from Single-CU to Multi-CU reduces total energy by 5.7\% for the MILP (276.9 to 261.2~kWh) and by 6.0\% for the heuristic (304.6 to 286.4~kWh).
Multi-CU operation also lowers peak hourly energy for both methods (MILP: 19.2 to 18.1~kWh; heuristic: 21.1 to 19.9~kWh), indicating more stable energy consumption under flexible DU-to-CU-UP association.
Thus, the Multi-CU advantage is retained with a comparable magnitude under the heuristic approximation.

The proposed heuristic also outperforms all cloud-placement baselines in total energy.
Relative to the 286.4~kWh Multi-CU heuristic result, total energy is 10.4\% higher for BF (316.1~kWh), 11.4\% higher for FFD (319.0~kWh), 12.4\% higher for MBFD (321.8~kWh), and 37.5\% higher for WF (393.8~kWh).
These differences arise because standard packing rules optimize local placement criteria rather than jointly considering consolidation, delay feasibility, and migration cost over time.
In particular, WF distributes the workload across all 12 servers and therefore misses most idle-server shutdown opportunities, whereas FFD and MBFD introduce substantially more migration churn.

\begin{figure}[pos=ht]
  \centering
  \includegraphics[width=0.9\linewidth]{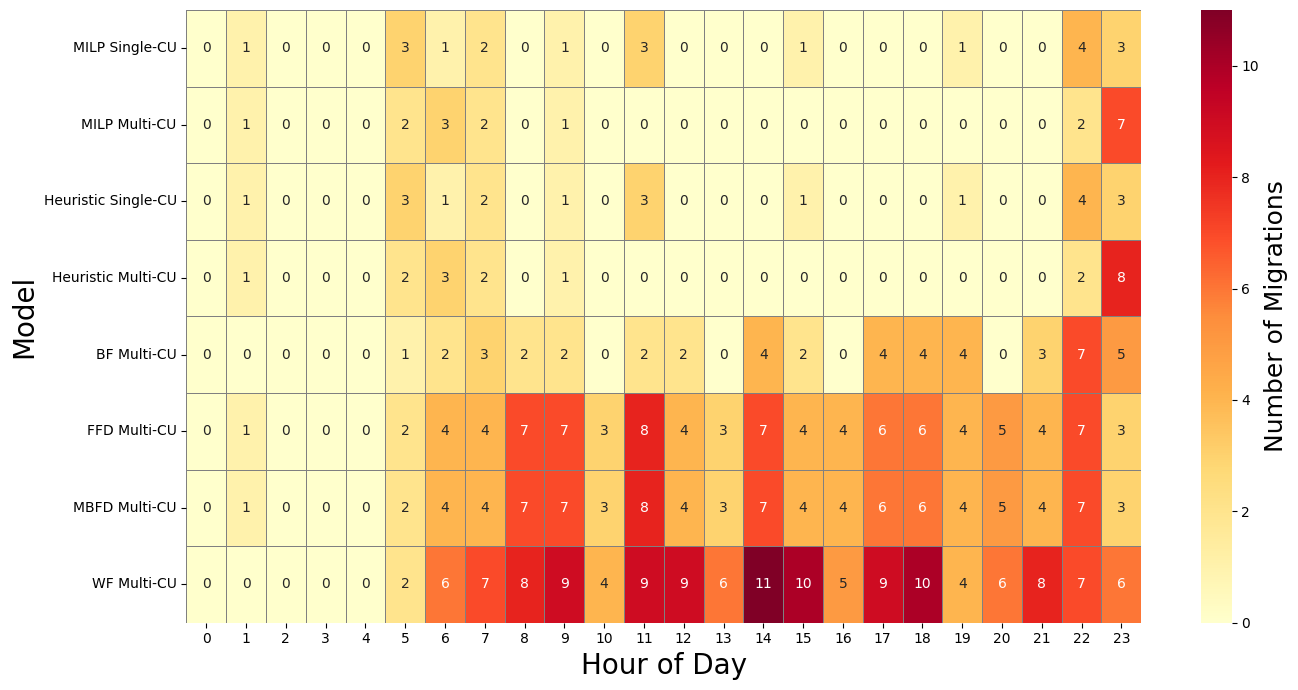}
  \caption{Total number of migrations over the 24-hour workload for the MILP, proposed heuristic, and cloud-placement baselines.}
  \label{fig:heuristic_number_of_migration}
\end{figure}

In addition, Fig.~\ref{fig:heuristic_number_of_migration} confirms that the heuristic closely follows the MILP's selective migration behaviour.
All methods inherit the previous placement at the beginning of each hour, so the reported migrations are not restart artifacts.
However, the static packing baselines do not explicitly penalize relocating an existing DU/CU-UP instance and may therefore repack instances as demand changes.
Consequently, in the Multi-CU scenario, the heuristic performs only 19 migrations, close to the MILP's 18, whereas BF, FFD, MBFD, and WF require 47, 93, 93, and 136 migrations, respectively.
The heuristic therefore retains MILP-like placement continuity while using migration selectively to preserve feasibility and enable consolidation.

\begin{figure}[pos=ht]
  \centering
  \includegraphics[width=1\linewidth]{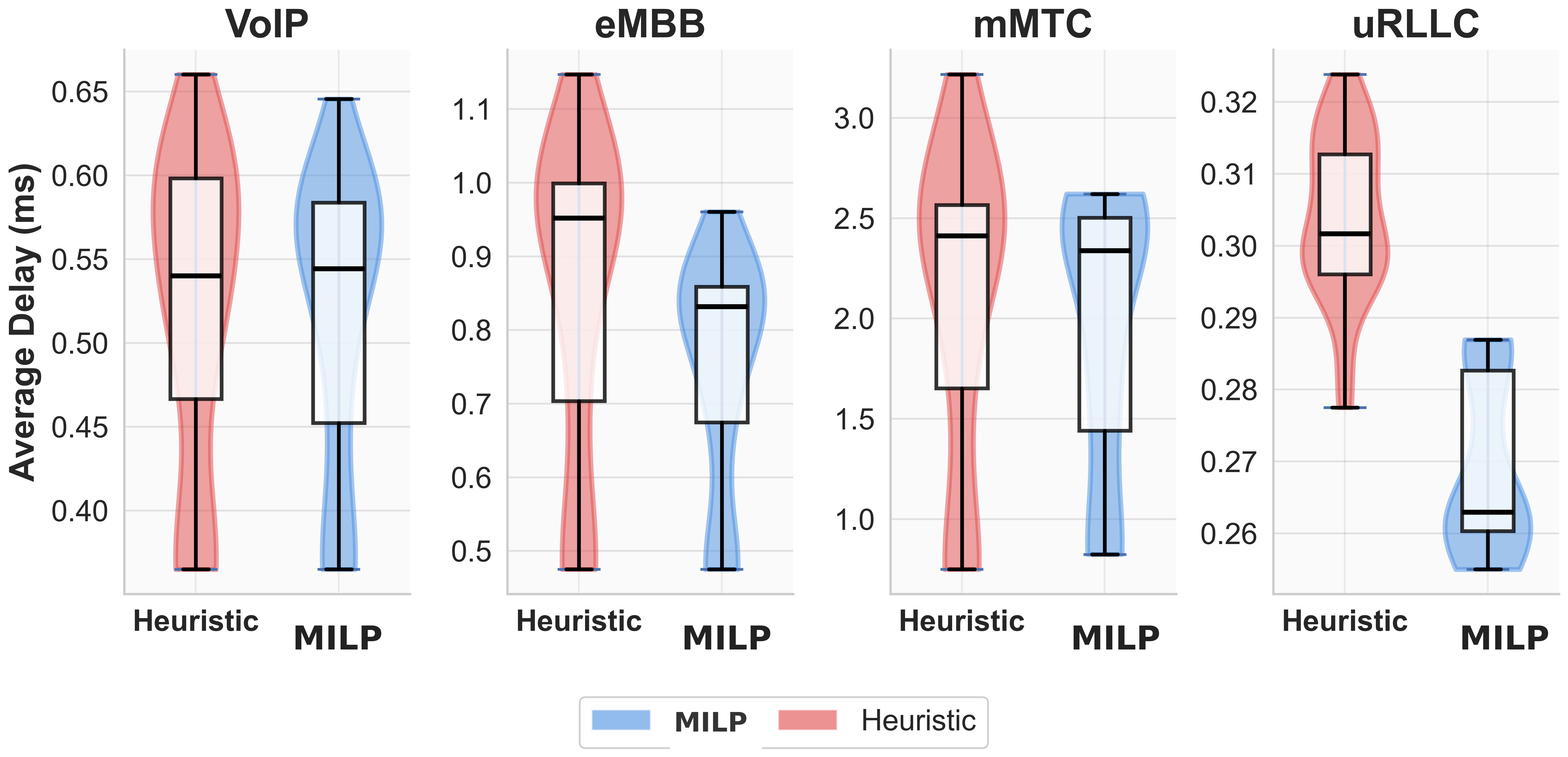}
  \caption{Comparison of slice-average one-way F1-U delay distributions.}
  \label{fig:combined_delay_results}
\end{figure}

Transmission energy is omitted from this comparison because it is negligible relative to placement and migration energy and follows the same pattern as the MILP results in Fig.~\ref{fig:ilp_transmission_energy}.
Finally, Fig.~\ref{fig:combined_delay_results} shows that the heuristic produces slice-average delay distributions close to those of the MILP.
Taken together, these results indicate that the heuristic approaches the MILP's energy and delay performance while reducing total energy and migration churn relative to the cloud-placement baselines.


\section{Conclusion and future work}

We introduced \PbName, a mixed-integer linear formulation for joint DU/CU-UP CNF placement and migration that minimizes server, transmission, wake-up, and migration energy under resource and one-way F1-U delay constraints while enabling the evaluation of flexible DU-to-CU-UP association strategies. We also developed \NameHeuristic, a deterministic four-phase $k$-means-based heuristic for logical pairing, physical mapping, consolidation, and feasibility repair.
In the evaluated 24-hour trace, Multi-CU reduces MILP energy by 5.7\% relative to Single-CU and requires fewer migrations during peak traffic; both scenarios retain one fixed CU-CP association per DU. The heuristic remains within 9.7\% of the Multi-CU MILP result and achieves a comparable Multi-CU energy reduction of 6.0\%. Its bounded clustering and repair procedures provide a polynomial-time alternative for larger instances, although the asymptotic analysis does not replace empirical runtime evaluation.

Future work will evaluate runtime scalability and integrate traffic forecasting to extend the current interval-by-interval model with predictive placement and migration decisions.

\section*{Acknowledgments}

We would like to thank Adel Larabi of Ericsson Montréal for his helpful discussions and verification efforts.
This work was supported by the Natural Sciences and Engineering Research Council of Canada (NSERC) and InnovÉÉ (INNOV-R Program) through the partnership with Ericsson and Environment and Climate Change Canada (ECCC) under Project ALLRP 566589-21 (grant identifier).

\bibliographystyle{elsarticle-num}
\bibliography{cas-refs}

@techreport{3gpp38401,
  author={{3GPP}},
  title={{NG-RAN}; Architecture Description},
  institution={3rd Generation Partnership Project (3GPP)},
  type={Technical Specification},
  number={TS 38.401},
  year={2026},
  month=apr,
  note={Version 19.2.0, Release 19}
}

@ARTICLE{7393804,
  author={Wang, Kezhi and Yang, Kun and Magurawalage, Chathura Sarathchandra},
  journal={IEEE Transactions on Cloud Computing}, 
  title={Joint Energy Minimization and Resource Allocation in C-RAN with Mobile Cloud}, 
  year={2018},
  volume={6},
  number={3},
  pages={760-770},
  doi={10.1109/TCC.2016.2522439}}

@inproceedings{agarwal2018joint,
  title={Joint VNF placement and CPU allocation in 5G},
  author={Agarwal, Satyam and Malandrino, Francesco and Chiasserini, Carla-Fabiana and De, Swades},
  booktitle={IEEE INFOCOM 2018-IEEE conference on computer communications},
  pages={1943--1951},
  year={2018},
  organization={IEEE}
}

@article{yang2021delay,
  title={Delay-Aware Virtual Network Function Placement and Routing in Edge Clouds},
  author={Yang, Song and Li, Fan and Trajanovski, Stojan and Chen, Xu and Wang, Yu and Fu, Xiaoming},
  journal={IEEE Transactions on Mobile Computing},
  volume={20},
  number={2},
  pages={445--459},
  year={2021},
  doi={10.1109/TMC.2019.2942306},
  publisher={IEEE}
}

@ARTICLE{Bo_2024,
  author={Yi, Bo and Wang, Jiacheng and He, Qiang and Wang, Xingwei and Huang, Min and Das, Sajal k. and Li, Keqin},
  journal={IEEE Transactions on Services Computing}, 
  title={Traffic Prediction-Based VNF Auto-Scaling and Deployment Mechanism for Flexible and Elastic Service Provision}, 
  year={2024},
  volume={17},
  number={5},
  pages={2959-2973},
}

@article{calheiros2011cloudsim,
  title={CloudSim: a toolkit for modeling and simulation of cloud computing environments and evaluation of resource provisioning algorithms},
  author={Calheiros, Rodrigo N and Ranjan, Rajiv and Beloglazov, Anton and De Rose, C{\'e}sar AF and Buyya, Rajkumar},
  journal={Software: Practice and experience},
  volume={41},
  number={1},
  pages={23--50},
  year={2011},
  publisher={Wiley Online Library}
}

@article{dai2024ran,
    author={Dai, Jiongyu and Li, Lianjun and Safavinejad, Ramin and Mahboob, Shadab and Chen, Hao and Ratnam, Vishnu V and Wang, Haining and Zhang, Jianzhong and Liu, Lingjia},
  journal={IEEE Transactions on Mobile Computing}, 
  title={O-RAN-Enabled Intelligent Network Slicing to Meet Service-Level Agreement (SLA)}, 
  year={2025},
  volume={24},
  number={2},
  pages={890-906},
  doi={10.1109/TMC.2024.3476338}
}

@inproceedings{delgado2022network,
  title={A network simulator for 5G virtualized networks},
  author={Delgado, Oscar and Jaumard, Brigitte and Ding, Zhiyi and Bishay, Fadi and Bissonnette, Vincent},
  booktitle={2022 IEEE 8th International Conference on Network Softwarization (NetSoft)},
  pages={237--239},
  year={2022},
  organization={IEEE}
}

@article{di2022optimization,
  title={Optimization over time of reliable 5G-RAN with network function migrations},
  author={Di Cicco, Nicola and Tonini, Federico and Cacchiani, Valentina and Raffaelli, Carla},
  journal={Computer Networks},
  volume={215},
  pages={109216},
  year={2022},
  publisher={Elsevier}
}

@article{dhahbi2021load,
  title={Load balancing in cloud computing using worst-fit bin-stretching},
  author={Dhahbi, Sami and Berrima, Mouhebeddine and Al-Yarimi, Fuad A. M.},
  journal={Cluster Computing},
  volume={24},
  number={4},
  pages={2867--2881},
  year={2021},
  publisher={Springer},
  doi={10.1007/s10586-021-03302-7}
}

@article{golkarifard2021dynamic,
  title={Dynamic VNF placement, resource allocation and traffic routing in 5G},
  author={Golkarifard, Morteza and Chiasserini, Carla Fabiana and Malandrino, Francesco and Movaghar, Ali},
  journal={Computer Networks},
  volume={188},
  pages={107830},
  year={2021},
  publisher={Elsevier}
}

@inproceedings{hojeij2023dynamic,
  title={Dynamic placement of O-CU and O-DU functionalities in open-ran architecture},
  author={Hojeij, Hiba and Sharara, Mahdi and Hoteit, Sahar and V{\`e}que, V{\'e}ronique},
  booktitle={2023 20th Annual IEEE International Conference on Sensing, Communication, and Networking (SECON)},
  pages={330--338},
  year={2023},
  organization={IEEE}
}

@article{hojeij2024flexible,
  title={On flexible association and placement in disaggregated RAN designs},
  author={Hojeij, Hiba and Ricardo, Guilherme Iecker and Sharara, Mahdi and Secci, Stefano},
  journal={Computer Communications},
  volume={238},
  pages={108166},
  year={2025},
  publisher={Elsevier},
  doi={10.1016/j.comcom.2025.108166}
}

@inproceedings{hojeij2025energy,
  title={Energy-Efficient Placement and Association in Disaggregated O-RAN},
  author={Hojeij, Hiba and Hoteit, Sahar and V\`eque, V\'eronique and Aravanis, Alexis I.},
  booktitle={2025 21st International Conference on Network and Service Management (CNSM)},
  pages={1--5},
  year={2025},
  doi={10.23919/CNSM67658.2025.11297540}
}

@article{ismail2024powergen,
  title={PowerGen: resources utilization and power consumption data generation framework for energy prediction in edge and cloud computing},
  author={Ismail, Leila and Materwala, Huned},
  journal={Procedia Computer Science},
  volume={238},
  pages={385--395},
  year={2024},
  publisher={Elsevier}
}

@article{klinkowski2020flowallocation,
  author={Klinkowski, Miroslaw},
  title={Optimization of Latency-Aware Flow Allocation in {NGFI} Networks},
  journal={Computer Communications},
  year={2020},
  volume={161},
  pages={344--359},
  doi={10.1016/j.comcom.2020.07.044}
}

@article{lin2020taxonomy,
  title={A taxonomy and survey of power models and power modeling for cloud servers},
  author={Lin, Weiwei and Shi, Fang and Wu, Wentai and Li, Keqin and Wu, Guangxin and Mohammed, Al-Alas},
  journal={ACM Computing Surveys (CSUR)},
  volume={53},
  number={5},
  pages={1--41},
  year={2020},
  publisher={ACM New York, NY, USA}
}

@article{moges2019energy,
  title={Energy-aware VM placement algorithms for the OpenStack Neat consolidation framework},
  author={Moges, Fikru Feleke and Abebe, Surafel Lemma},
  journal={Journal of Cloud Computing},
  volume={8},
  number={1},
  pages={2},
  year={2019},
  publisher={Springer}
}

@article{murti2024reconfigurations,
  title={Deep Reinforcement Learning for Orchestrating Cost-Aware Reconfigurations of vRANs},
  author={Murti, Fahri Wisnu and Ali, Samad and Iosifidis, George and Latva-aho, Matti},
  journal={IEEE Transactions on Network and Service Management},
  volume={21},
  number={1},
  pages={200--216},
  year={2024},
  doi={10.1109/TNSM.2023.3292713},
  publisher={IEEE}
}

@article{monaco2026crown,
  title={{CROWN}: Cross-attention reinforcement learning for O-RAN wireless networks},
  author={Monaco, Doriana and Sacco, Alessio and Marchetto, Guido},
  journal={Computer Networks},
  volume={282},
  pages={112276},
  year={2026},
  doi={10.1016/j.comnet.2026.112276}
}

@article{municio2023ran,
  title={O-ran: Analysis of latency-critical interfaces and overview of time sensitive networking solutions},
  author={Municio, Esteban and Garcia-Aviles, Gines and Garcia-Saavedra, Andres and Costa-P{\'e}rez, Xavier},
  journal={IEEE Communications Standards Magazine},
  volume={7},
  number={3},
  pages={82--89},
  year={2023},
  publisher={IEEE}
}

@inproceedings{mushtaq2023optimal,
  title={Optimal functional splitting, placement and routing for isolation-aware network slicing in NG-RAN},
  author={Mushtaq, Maria and Golkarifard, Morteza and Shahriar, Nashid and Boutaba, Raouf and Saleh, Aladdin},
  booktitle={2023 19th International Conference on Network and Service Management (CNSM)},
  pages={1--5},
  year={2023},
  organization={IEEE},
  doi={10.23919/CNSM59352.2023.10327830}
}

@article{pires2025optimizing,
  title={Optimizing Energy Consumption for vRAN Placement in O-RAN Systems With Flexible Transport Networks},
  author={Pires Jr., William Teixeira and Almeida, Gabriel Matheus Faria de and Corr{\^e}a, Sand Luz and Both, Cristiano Bonato and Pinto, Leizer de Lima and Cardoso, Kleber Vieira},
  journal={IEEE Open Journal of the Communications Society},
  volume={6},
  pages={4279--4294},
  year={2025},
  doi={10.1109/OJCOMS.2025.3568689}
}

@article{ramanathan2021live,
  title={Live migration of virtual machine and container based mobile core network components: A comprehensive study},
  author={Ramanathan, Shunmugapriya and Kondepu, Koteswararao and Razo, Miguel and Tacca, Marco and Valcarenghi, Luca and Fumagalli, Andrea},
  journal={IEEE Access},
  volume={9},
  pages={105082--105100},
  year={2021},
  publisher={IEEE}
}

@article{ruiz2020genetic,
  title={Genetic algorithm for holistic VNF-mapping and virtual topology design},
  author={Ruiz, Lidia and Barroso, Ram{\'o}n J Dur{\'a}n and De Miguel, Ignacio and Merayo, Noem{\'\i} and Aguado, Juan Carlos and De La Rosa, Ramon and Fern{\'a}ndez, Patricia and Lorenzo, Rub{\'e}n M and Abril, Evaristo J},
  journal={IEEE Access},
  volume={8},
  pages={55893--55904},
  year={2020},
  publisher={IEEE}
}

@article{sahin2026rfdr,
  title={{RFD-R}: AI-driven dynamic repacking framework for cloud-native O-RAN functions},
  author={\c{S}ahin, Cihan and G{\"u}n, Mahmut},
  journal={Computer Networks},
  volume={282},
  pages={112265},
  year={2026},
  doi={10.1016/j.comnet.2026.112265}
}

@article{sen2025slice,
  title={Slice aware baseband function splitting and placement in disaggregated 5G Radio Access Network},
  author={Sen, Nabhasmita and A., Antony Franklin},
  journal={Computer Networks},
  volume={257},
  pages={110908},
  year={2025},
  publisher={Elsevier},
  doi={10.1016/j.comnet.2024.110908}
}

@article{shirvastava2017best,
  title={Best fit based VM allocation for cloud resource allocation},
  author={Shirvastava, Saurabh and Dubey, Rahul and Shrivastava, Manish},
  journal={International Journal of Computer Applications},
  volume={158},
  number={9},
  year={2017},
  publisher={Foundation of Computer Science}
}

@article{subramanya2021predictive,
  author={Subramanya, Tejas and Riggio, Roberto},
  journal={IEEE Transactions on Network and Service Management},
  title={Centralized and Federated Learning for Predictive {VNF} Autoscaling in Multi-Domain {5G} Networks and Beyond},
  year={2021},
  volume={18},
  number={1},
  pages={63--78},
  doi={10.1109/TNSM.2021.3050955}
}

@article{tang2019dynamic,
  author={Tang, Lun and He, Xiaoyu and Zhao, Peipei and Zhao, Guofan and Zhou, Yu and Chen, Qianbin},
  title={Virtual Network Function Migration Based on Dynamic Resource Requirements Prediction},
  journal={IEEE Access},
  volume={7},
  pages={112348--112362},
  year={2019},
  doi={10.1109/ACCESS.2019.2935014},
  publisher={IEEE}
}

@article{takci2025data,
  title={Data centres as a source of flexibility for power systems},
  author={Takci, Mehmet T{\"u}rker and Qadrdan, Meysam and Summers, Jon and Gustafsson, Jonas},
  journal={Energy Reports},
  volume={13},
  pages={3661--3671},
  year={2025},
  publisher={Elsevier}
}

@article{tran2025proactive,
  title={Proactive Service Assurance in 5G and B5G Networks: A Closed-Loop Algorithm for End-to-End Network Slices},
  author={Tran, Nguyen Phuc and Delgado, Oscar and Jaumard, Brigitte},
  journal={IEEE Transactions on Network and Service Management},
  volume={23},
  pages={668--680},
  year={2025},
  publisher={IEEE}
}

@techreport{Trojer2021PacketFronthaul,
  author = {Trojer, Elmar and Skogman, Viktor and Olsson, Andreas and Thyni, Tomas and Forsman, Mats and {\"O}sterling, Jacob and Cederholm, Daniel and Berg, Miguel and Stake, Roger},
  title = {Packet fronthaul -- design choices towards versatile {RAN} deployments},
  institution = {Ericsson},
  year = {2021},
  type = {White Paper},
  url = {https://www.ericsson.com/en/reports-and-papers/white-papers/packet-fronthaul-design-choices}
}

@article{tseliou2019netslic,
  author={Tseliou, Georgia and Adelantado, Ferran and Verikoukis, Christos},
  title={{NetSliC}: Base Station Agnostic Framework for Network Slicing},
  journal={IEEE Transactions on Vehicular Technology},
  year={2019},
  volume={68},
  number={4},
  pages={3820--3832},
  doi={10.1109/TVT.2019.2902320}
}

@article{verma2024vnfscaling,
  author={Verma, Rahul and Sivalingam, Krishna M.},
  journal={IEEE Access},
  title={Design and Analysis of {VNF} Scaling Mechanisms for {5G}-and-Beyond Networks Using Federated Learning},
  year={2024},
  volume={12},
  pages={129826--129843},
  doi={10.1109/ACCESS.2024.3458437}
}

@article{zorello2022baseband,
  author={Zorello, Ligia Maria Moreira and Bliek, Laurens and Troia, Sebastian and Guns, Tias and Verwer, Sicco and Maier, Guido},
  title={Baseband-Function Placement With Multi-Task Traffic Prediction for {5G} Radio Access Networks},
  journal={IEEE Transactions on Network and Service Management},
  year={2022},
  volume={19},
  number={4},
  pages={5104--5119},
  doi={10.1109/TNSM.2022.3190059}
}

\begin{biography}[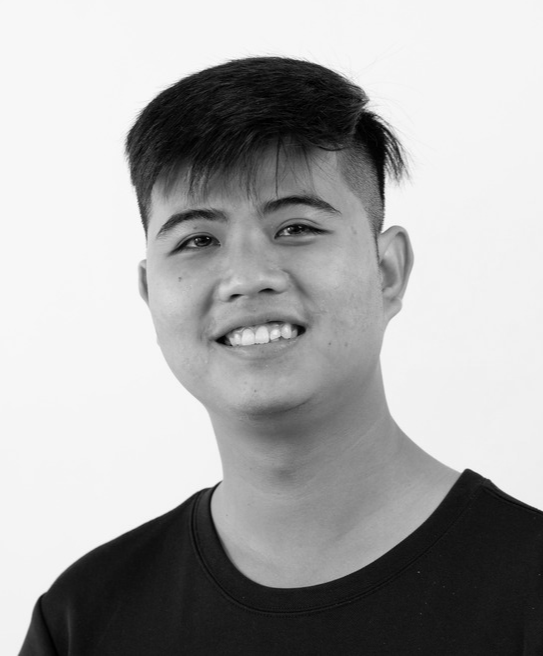]{Nguyen Phuc Tran}
  received the M.S. degree in Computer Science from the University of Information Technology, Vietnam National University Ho Chi Minh City, Vietnam, in 2020. He is currently pursuing the Ph.D. degree at Concordia University, Montreal, Quebec, Canada. He has more than five years of industry experience as a senior software engineer in systems development and telecommunications, with expertise in system optimization, security, quality assurance, data analytics, technical leadership, and stakeholder engagement. His research interests include artificial intelligence, particularly Machine Learning, Reinforcement Learning and large language models, for mobile communication networks, with an emphasis on resource allocation, energy-efficient and sustainable networking, system design, root-cause analysis, and network optimization.
\end{biography}

\begin{biography}[Figures/authors/brigitte.pdf]{Brigitte Jaumard}
  (Senior Member, IEEE) is the scientific director of Confiance IA (Intelligence Artificielle), an industrial research consortium on trustworthy AI supported by the Quebec government. She is also a professor in the Computer Science and Software Engineering Department at Concordia University. Her research focuses on mathematical modelling and algorithm design, including large-scale optimization and machine learning, for problems arising in communication, transportation, and logistics networks. Recent studies include efficient optimization and machine-learning algorithms for network design, dimensioning and provisioning, scheduling in edge computing and clouds, and 5G networks. During her 2020--2021 sabbatical year, she was a senior advisor for the Montreal Ericsson Global Artificial Intelligence Accelerator research center and the chief scientist of the Centre de Recherche Informatique de Montréal. She was ranked among the top 2\% of scientists in her field according to a 2021 citation-based study and held Tier I Canada and Concordia Research Chairs during 2000--2019. She has published more than 300 papers in operations research and telecommunications.
\end{biography}

\begin{biography}[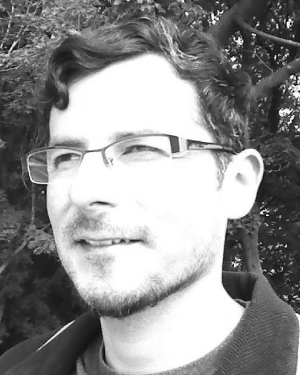]{Oscar Delgado}
  received the Ph.D. degree in electrical engineering from McGill University, Montreal, in 2016. After his Ph.D., he was a postdoctoral researcher at the Telecommunications and Signal Processing Laboratory, Department of Electrical and Computer Engineering, McGill University. He is currently a research associate at \'{E}cole de Technologie Sup\'{e}rieure. His research interests include applications of 5G wireless mobile communication technologies, Artificial Intelligence and Machine Learning, network virtualization, service assurance, green wireless systems, video-traffic management, resource-allocation strategies, and energy-efficient algorithms.
\end{biography}

\end{document}